\documentclass[aps,prd,floatfix,superscriptaddress]{revtex4} 
\usepackage{amssymb,amsmath,amsfonts,latexsym,graphicx,epsfig}
\usepackage{float}
\usepackage{titlesec}
\usepackage{color}
\newcommand{\eq}{\begin{eqnarray}}
\newcommand{\en}{\end{eqnarray}}
\titleformat{\section}{\large\bfseries}{\thesection}{1em}{}

\begin{document}

\title{Radiative transitions of charmonium states in \\ 
the covariant confined quark model} 

\author{Gurjav Ganbold}
\affiliation{Bogoliubov Laboratory of Theoretical Physics,
Joint Institute for Nuclear Research, 141980 Dubna, Russia}
\affiliation{Institute of Physics and Technology, Mongolian Academy
of Sciences, Enkh Taivan 54b, 13330 Ulaanbaatar, Mongolia}
\author{Thomas Gutsche}
\affiliation{Institut f\"ur Theoretische Physik,
Universit\"at T\"ubingen,
Kepler Center for Astro and Particle Physics,
Auf der Morgenstelle 14, D-72076 T\"ubingen, Germany}
\author{Mikhail A. Ivanov} 
\affiliation{Bogoliubov Laboratory of Theoretical Physics,
Joint Institute for Nuclear Research, 141980 Dubna, Russia}
\author{Valery E. Lyubovitskij}
\affiliation{Institut f\"ur Theoretische Physik,
Universit\"at T\"ubingen,
Kepler Center for Astro and Particle Physics,
Auf der Morgenstelle 14, D-72076 T\"ubingen, Germany}
\affiliation{Departamento de F\'\i sica y Centro Cient\'\i fico
Tecnol\'ogico de Valpara\'\i so-CCTVal, Universidad T\'ecnica
Federico Santa Mar\'\i a, Casilla 110-V, Valpara\'\i so, Chile}
\affiliation{Millennium Institute for Subatomic Physics at the High-Energy Frontier (SAPHIR) of ANID, \\
Fern\'andez Concha 700, Santiago, Chile}
\affiliation{Department of Physics, Tomsk State University,
634050 Tomsk, Russia}
\affiliation{Tomsk Polytechnic University, 634050 Tomsk, Russia}

\begin{abstract}
We have studied the dominant radiative transitions of the charmonium $S$-
and $P$-wave states within the covariant confined quark model. 
The gauge invariant leading-order
transition amplitudes have been expressed by using either the conventional
Lorentz structures, or the helicity amplitudes, where it was effective.
The renormalization couplings of the charmonium states
have been strictly fixed by the compositeness conditions that excludes the
constituent degrees of freedom from the space of physical states.  We use the
basic model parameters for the constituent c-quark mass $m_c=1.80$ GeV 
and the global infrared cutoff $\lambda=0.181$ GeV. 
We additionally introduce only one adjustable parameter $\varrho>0$ common
for the charmonium states $\eta_c({}^1\!S_0)$, $J/\psi({}^3\!S_1)$,
$\chi_{c0}({^{3}}\!P_{0})$, $\chi_{c1}({^{3}}\!P_{1})$,  $h_c({^{1}}\!P_{1})$, 
and $\chi_{c2}({^{3}}\!P_{2})$ to describe the quark distribution inside the
hadron. This parameter describes the ratio between the charmonium ``size'' and
its physical mass. The optimal value $\varrho=0.485$ has been fixed by fitting
the latest data for the partial widths of the one-photon radiative decays of the
triplet $\chi_{cJ}({^{3}}\!P_{J}),~J=\{0,1,2\}$. Then, we calculate corresponding
fractional widths for states $J/\psi({}^3\!S_1)$ and $h_c({^{1}}\!P_{1})$.
Estimated results are in good agreement with the latest data.
By using the fraction data from PDG2020 and our estimated partial decay
width for $h_c({^{1}}\!P_{1})$ we recalculate the ``theoretical full width''
$\Gamma^{\rm theor}_{h_c}  \simeq ( 0.57 \pm 0.12 )$ MeV in comparison
with latest data $\Gamma^{\rm exp}_{h_c} \simeq (0.7\pm 0.4)$ MeV.
We also repeated our calculations by gradually decreasing the global cutoff
parameter and revealed that the results do not change for any
$\lambda<0.181$ GeV up to the ``deconfinement'' limit.
\end{abstract}

\maketitle

\section{Introduction}
\label{sec:intro}

Charmonium is a bound state of a charm quark and antiquark.
The properties of charmonium states have been intensively studied within various
theoretical frameworks based on and motivated by QCD since the first
charmonium state $J/\psi$ was observed in 1974~\cite{Aubert:1974js,Augustin:1974xw}.
The charmonium states are unusual since the quark masses are much larger
than the typical confinement scale and they have low-lying excited states
observed in different experiments~\cite{LHCb15Cleo13}.
Typically, low-lying $c{\bar c}$ mesons have narrow widths and their dominant
radiative transitions are one-photon decay modes.  Particularly, the decays of the
charmonium states  $J/\psi$, $\chi_{c0}$, $\chi_{c1}$, $h_c$, and $\chi_{c2}$,
which are below the $D{\bar D}$ threshold, have been observed and measured fairly accurately.

The state $h_c$ escaped experimental detection for a long time. Only
many years later CLEO succeeded to isolate this state~\cite{Rosner:2005ry}
and observed that its prominent mode is $h_c(1P)\to \gamma \eta$.
The branching fraction was later accurately measured
at the BESIII experiment~\cite{Ablikim:2010rc}.
Recent studies of $\chi_{c0}$, $\chi_{c1}$, and $\chi_{c2}$ mesons at hadron
colliders have exploited the radiative decays
$\chi_{c0,c1,c2}\to\gamma J/\psi$~\cite{LHCb-PAPER-2011-019,%
LHCb-PAPER-2013-028,aaij17} and have found that the branching fractions
are fairly large, allowing us to detect a signal despite the high background.

In the description of hadron structure within the Standard Model, many theories are based on
phenomenological models and effective theories in varying degrees of sophistication
in their relation to the underlying QCD.
The charmonium system presents itself as an ideal testing ground to explore the validity 
of basic model assumptions
due to its small binding energy which allows for perturbative calculations,
even by means of nonrelativistic approaches.

Radiative decays in  charmonium play an important role in the understanding
of their structure and can serve as a testing ground for a number of theories and
models  (see, e.g., Refs.~\cite{Barnes:2005pb,Voloshin:2007dx}).
Properties of the electromagnetic (radiative) decays of low-lying charmonium
states can be estimated with the widely used quark potential
models~\cite{Godfrey:1985xj}-\cite{weijun2016}, 
but these models are insufficient in obtaining high precision.
Besides the conventional potential models, other approaches such as lattice simulations
QCD~\cite{Dudek:2006ej,Dudek:2009kk,Chen:2011kpa,Becirevic:2012dc},
QCD sum rules~\cite{Khodjamirian:1979fa,Beilin:1984pf,Zhu:1998ih},
effective Lagrangian approaches~\cite{DeFazio:2008xq,Wang:2011rt},
nonrelativistic effective field theories of QCD~\cite{Brambilla:2005zw,%
Brambilla:2012be,Pineda:2013lta}, quark models~\cite{Ebert:2003,brus2020},
approaches based on solutions of the Bethe-Salpeter equations~\cite{Wang:2010ej},
light-front quark model~\cite{Ke:2013zs}, and Coulomb gauge approach~\cite{Guo:2014zva}
have been employed to deal with the theoretical description of
radiative decays of charmonium.

For the radiative transitions of the low-lying charmonium states
discrepancies still exist between the theoretical predictions and experimental
measurements.
Particularly, calculations with the nonrelativistic potential model~\cite{Swanson:2005} 
and in the Coulomb gauge approach~\cite{Guo:2014zva} result in
large widths $\Gamma(J/\psi\to\gamma\eta_c(1S) )\simeq 2.9$~keV,
about a factor of 2 larger than the latest world average of data
reported by the PDG~\cite{PDG20}. Quark models fail to reproduce the measured branching width
$\Gamma(J/\psi \to\gamma\eta_c)$ and, instead, obtain a significantly larger
value~\cite{Eichten:2007qx,Voloshin:2007dx,Swanson:2005}.
Recently, the electromagnetic transitions of charmonium states have been
studied with a constituent quark model and a reasonable description of
the radiative transitions of the well-established charmonium states $J/\psi$,
$\psi(2S)$, $\chi_{cJ}(1P)$, $h_c(1P)$, and $\psi(3770)$ has been
obtained~\cite{weijun2016},
but the numerical results differ from the worldwide data.
Some studies on the radiative transition properties of $\chi_{c0,c1}(1P)$ were
carried out in Lattice QCD as well~\cite{Dudek:2006ej,Chen:2011kpa},
however, good descriptions are still not obtained due to technical restrictions.
Although some comparable predictions from different models have been obtained,
strong model dependencies still exist.

Recently, the Particle Data Group (PDG)~\cite{PDG20} has reported that the treatment
of the branching ratios of the orbital charmonium excitations has undergone an important
restructuring.  Hereby, the PDG used footnotes to indicate the branching ratios
actually given by the experiments and the quantities they use to derive them
from the true combination of branching ratios actually measured~\cite{PDG20}.
Meanwhile, in the wake of the recent measurements of charmonium production
in $b$-hadron decays  performed by the LHCb Collaboration~\cite{aaij17} a theoretical
analysis of the quasiparticle properties of the orbital charmonium excitations
serves a good testing ground and may contribute to a deeper understanding of
the underlying physical processes.

The construction of appropriate theoretical models to explain and implement
the confinement phenomenon is important in modern particle physics. 
Our relativistic approach has an advantage to describe different composite
systems by using the same theoretical principles. The approach is manifestly
Lorentz invariant, consistent with symmetries of QCD and electroweak theory and
consistently incorporates symmetry breaking. For the processes involving heavy
quarkonia the important limits like nonrelativistic and heavy quark limits are
straightforwardly  applied in our approach. In these sense our approach goes
beyond nonrelativistic potential model and is able to take into account
relativistic effects.  Therefore, our approach goes beyond nonrelativisitc
picture without spoiling ``classic'' results and taking into account
relativistic effects by using the quantum field approach. 

One of the key point of our approach is the fulfillment of Weinberg-Salam (WS)
framework of compositeness condition (see, Refs.~\cite{Salam:1962ap}-\cite{Efimov:1993ei}) 
to exclude totally any
unphysical degrees of freedom from consideration dealing with bound states.
This framework is consistent with all axioms of quantum field theory and has
clear nonrelativistic reduction proposed by Weinberg in Ref.~~\cite{Weinberg:1962hj}. 
Originally, Salam~\cite{Salam:1962ap} and Weinberg~\cite{Weinberg:1962hj} 
applied this approach for a description of deuteron as proton-neutron bound state.
Later on, our group extended the ideas of Salam and Weinberg to various
composite systems, starting from conventional hadrons (mesons and baryons)
[see, e.g., Refs.~\cite{Ivanov:1996pz}-\cite{Faessler:2003yf}] up to 
hybrids~\cite{Ivanov:1985zw}, tetraquarks~\cite{Dubnicka:2010kz,Goerke:2016hxf} 
pentaquarks~\cite{Gutsche:2019mkg}, and 
hadronic molecules~\cite{Faessler:2007gv}-\cite{Dong:2008gb}.  

Application of our framework for study of bound states is based on
the following steps:

(1) First, we derive a phenomenological Lagrangian, which is manifestly Lorenz
covariant and gauge invariant and describes the interaction of the bound state
and  its constituents via interpolating currents with respective
$J^{PC}$ quantum numbers;

(2) Second, the coupling constant of the bound state with the constituents is
fixed by solving the WS condition for vanishing of the renormalization
function of the bound state $Z = 1 - \Pi' = 0$ [1]-[4], where $\Pi'$ is
the derivative of the mass operator of the bound state induced by interaction
Lagrangian of the bound state  with its constituents. The WS condition leads
to zero probability of finding bare states among bound states.
In other words, the bound state is always dressed by their constituents;

(3) Third, we construct the $S$-matrix operator, which consistently generates
matrix elements of physical processes involving bound states. Matrix elements
are represented by a set of Feynman diagrams, which are calculated using
the methods of quantum field theory. The convergence of Feynman diagrams is
guaranteed by introducing the cutoff regularization  consistent with gauge
invariance. Loop integration in the integrals  corresponding to Feynman
diagrams is performed in Euclidean region using the Wick rotation in momenta.

In our recent studies (see details in Refs.~\cite{Ivanov:1996pz}-\cite{Gutsche:2019mkg})   
we developed different mechanisms of confinement
and applied them to the actual problems in particle physics. As result we proposed 
and developed a series of models united by the physical principles discussed above. 
Last decade we proposed and developed new covariant confined quark 
model (CCQM)~\cite{bran10ivan17}, which 
implements effective quark confinement with convolutions of local quark
propagators and vertex functions accompanied by an infrared cutoff of the scale
integration that prevents any singularities in matrix elements. Furthermore, the CCQM
has been applied to multiquark meson structure, form factors and angular decay
characteristics of light and heavy  baryons. Possible new physics effects in the
exclusive decays of $\bar{B}^0$  and transitions of $\Lambda_b$ have been
studied in some extension of the Standard Model by taking into account right-handed
vector (axial), left- and right-handed (pseudo)scalar, and tensor current
contributions. In the framework  of the CCQM we demonstrated the equivalence
of Yukawa-type and Fermi-type theories under two constraint equations
determining the corresponding couplings and meson masses. The Fermi coupling
was evaluated as a function of meson mass and we obtained a new smooth behavior
for the resulting curve~\cite{Ganbold:2014pua}. 
Inspired by recent measurements we studied the
radiative decays of charmonium states below the $D {\bar D}$ threshold in 
a relativistic quark model with analytic confinement.
We stress, that decays
of heavy quarkonia have been considered in our framework earlier in~\cite{Ivanov:2000aj}. 
More recently, we have studied
implications of new physics in the decays of the $B_c$ meson into final charmonium
states within the standard model and beyond. 
Properties of baryons with two and three heavy quarks have been
considered in Refs.[8].
In this report we present a theoretical investigation on the hadronic structure and
dominant radiative transitions of the charmonium excitations within
the CCQM~\cite{bran10ivan17}.

The paper is organized as follows.
A brief description of the CCQM is introduced in Sec.~II.
In Sec.~III we consider the renormalization couplings of hadrons which are
constrained by the compositeness condition and thereby guarantee the
absence of constituent degrees of freedom from the space of physical states.
The dominant one-photon radiative transitions of the charmonium $S$- and $P$-wave
states, their leading-order transition amplitudes are studied in Sec.~IV.
In Sec.~V we present numerical results on the renormalized couplings and
fractional decay widths of charmonium states.
In Sec.~VI we go beyond the infrared confinement mode in the CCQM and repeat
our calculations by gradually decreasing the fixed global cutoff parameter for
confinement ($\lambda=181~\mathrm{MeV}$) up to the deconfinement limit
corresponding to $\lambda\to 0$. In Sec.VII we discuss the obtained
results by comparing them with some recent theoretical predictions and
experimental data~\cite{PDG20}. A summary is given in Sec.~VIII.

\section{Approach}
\label{sec:model}

The covariant confined quark model (CCQM) is an effective quantum
field approach to hadronic physics, which is based on a relativistic
Lagrangian describing the interaction of a hadron with its constituent quarks. 
Detailed descriptions of the model, as well as the calculational techniques
used for the quark-loop evaluation may be found
in~\cite{bran10ivan17,Gutsche:2018nks,%                                                                          
Dubnicka:2015iwg, Gutsche:2015mxa,Gutsche:2013pp,Ivanov:2015woa}
and references therein.
The CCQM  represents an appropriate theoretical framework to perform an
analysis of the recent measurement of radiative transitions involving the
charmonium ground ($J/\psi$) and excited states
($\chi_{c0}, \chi_{c1},h_c,\chi_{c2}$) reported
by the LHCb Collaboration~\cite{aaij17}.

The hadron is described by a field $H (x)$, which satisfies
the corresponding equation of motion, while the quark part is introduced
by an interpolating quark current $J_{H}(x)$ with the hadron quantum
numbers. In the case of mesons, the interaction Lagrangian reads
\eq
{\cal L}_{int}  & = & g_{H}\cdot H(x)\cdot
J_{H}(x)+{\rm H.c.}  \,, \nonumber\\
J_{H}(x) & = & \int\!\! dx_{1}\!\! \int\!\! dx_{2}
F_{H}(x,x_1,x_2)\,\bar{q}_2(x)
\Gamma_H\,q_1(x) \,,
\label{Lagrangian}
\en
where  $\Gamma_{H}$ is the Dirac matrix ensuring the quantum numbers
of the meson. The quark-meson coupling $g_{H}$ is determined by the
so-called compositeness condition, which imposes that the wave function
renormalization constant of the hadron has to be equal to zero: $Z_H = 0$.

The vertex function
\eq
F_H(x,x_1,x_2)
=  \delta^{(4)} (x - w_1 x_1 - w_2 x_2) \, \Phi_H\Big[ (x_1 - x_2 )^2\Big] \,,
\en
effectively describes the distribution of constituent quarks
inside the meson, where $w_i = m_i/(m_1+m_2),\: i=1,2$ is the fraction
of quark masses with $w_1+w_2=1$.

The Fourier transform of the translationally invariant vertex function
$\Phi_{H}\Big[\left(x_{1}-x_{2}\right)^{2}\Big]$ in momentum space
is required to fall off in the Euclidean region in order to provide the
ultraviolet convergence of the loop integrals. We use a simple
Gaussian form written as
\eq
\widetilde{\Phi}_H\left(-p^2\right)
= \exp\left( s_H \!\cdot\! p^2  \right), \qquad   s_H \equiv 1 / \Lambda_H^2 \,,
\label{vertex}
\en
where $\Lambda_H$ is an adjustable hadron size-related parameter of the CCQM.

We note, any choice for  $\Phi_H$ is appropriate as long as it falls off
sufficiently fast in the ultraviolet region to render the corresponding
Feynman diagrams ultraviolet finite.

For the quark propagator we use the Fock-Schwinger representation:
\eq
\widetilde{S}(\hat{k}) =  \left( m + \hat{k} \right) \intop_{0}^{\infty}
d\alpha\: \exp\big(-\alpha \left(m^2 - k^2 \right) \big) \,.
\label{propagator}
\en
Further on, we consider the hadron mass operator, matrix elements of hadronic
transitions are represented by quark-loop diagrams, which are described as
convolutions of the corresponding quark propagators and vertex functions.

In our previous papers we have shown that any hadronic matrix element
containing loops can be finally written in the form
$\Pi^{0} = N_c \int\limits_0^\infty\! d^n \alpha\, f(\alpha_1,\ldots,\alpha_n)$,
where $N_c=3$ is the number of colors and $f$ is the resulting integrand corresponding to a given
diagram.
It is convenient to convert the set of Fock-Schwinger parameters into
a simplex by adding the integral
$1=\int\limits_0^\infty\!dt\,\delta\big(t-\sum\limits_{i = 1}^n {{\alpha _i}}\big)$
as follows:
\eq
\Pi^{0} = N_c \int\limits_0^\infty\! dt \, t^{n-1} \int\limits_0^1\! d^n \alpha \,\,
\delta\Big(1 -  \sum\limits_{i=1}^n \alpha_i \Big)
f(t\alpha_1,\ldots,t\alpha_n) \,.
\label{integ0}
\en
The integral in Eq.~(\ref{integ0}) diverges for $t\to\infty$, if the kinematic
variables allow for the appearance of branch points corresponding to the
creation of free quarks. However, these possible threshold singularities
disappear if one cuts off the integral for large values of t:
\eq
\Pi^{\lambda} = N_c \int\limits_0^{1/\lambda^2}\! dt \, t^{n-1}
\int\limits_0^1\! d^n \alpha \,\,
\delta\Big(1 -  \sum\limits_{i=1}^n \alpha_i \Big)
f(t\alpha_1,\ldots,t\alpha_n) \,.
\label{integ1}
\en
The parameter $\lambda$  is the so-called infrared cutoff parameter,
which is introduced to avoid the appearance of a branch point
corresponding to the creation of free quarks, and taken to be universal
for all physical processes. The multidimensional integral in Eq.~(\ref{integ1})
is calculated numerically by using a \textsc{fortran} code.

The CCQM consists of several free parameters: the constituent quark
masses $m_q$, the hadron size parameters $\Lambda_H$, and the
universal infrared cutoff parameter $\lambda$.
The model parameters are determined by minimizing $\chi^2$ in
a fit to the latest available experimental data and some lattice results.
In doing so, we have observed that the errors of the fitted parameters are of
the order of  $\sim 10 \%$.

Particularly, the central values of the basic parameters updated
in Refs.~\cite{Ganbold:2014pua,Gutsche:2015mxa} are (in GeV):
\begin{equation}
\begin{tabular}{ c c c c  c  }
\quad $m_{u/d}$ \quad & \quad $m_s$ \quad & \quad $m_c$ \quad
& \quad $m_b$ \quad & \quad $\lambda$ \quad
\\\hline
 \quad 0.241 \quad & \quad 0.428 \quad & \quad 1.67 \quad &
\quad 5.05 \quad & \quad 0.181
\end{tabular}
\label{centralparameters}
\end{equation}

Obviously, the errors of our calculations within the CCQM are expected
to be about $\pm 10 \%$.
Below we apply the CCQM to estimate the decay widths of the dominant
(one-photon) radiative transitions of $\chi_{cJ}$ mesons.
The first step in the calculation of physical observables is the determination of the
couplings $g_{H}$ of the participating hadrons according to Eq.~(\ref{Lagrangian}).

\section{Renormalization couplings}
\label{sec:couplings}

\begin{figure}[H]
\begin{center}
\includegraphics[scale=0.5]{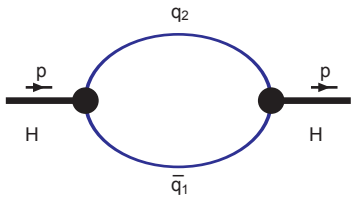}
\caption{\small Self-energy diagram for a meson.}
\label{fig1}
\end{center}
\end{figure}

According to the CCQM, the renormalization coupling $g_{H}$ of a hadron
in Eq.~(\ref{Lagrangian}) should be defined from the compositeness
condition as follows:
\eq
Z_H = 1 -  g_H^2 \tilde{\Pi}_H^{'} \!\! \left( M_H^2 \right) = 0 \,,
\qquad \tilde{\Pi}_H^{'} \!\! \left( p^2 \right) = \frac{d}{dp^2} \,
\tilde{\Pi}_H^{(1)}\!\! \left( p^2 \right) \,,
\label{gren}
\en
where $\tilde{\Pi}_H^{(1)}\!\! \left( p^2 \right)$ is the diagonal (scalar) part
of the hadron self-energy operator. For a meson the mass
operator corresponding to the self-energy diagram in Fig.~\ref{fig1} reads
\begin{eqnarray}
\tilde{\Pi}_{H}(p)
&=&  N_c \int\!\! \frac{dk}{(2\pi)^4i} \widetilde\Phi^2_H \left(-k^2\right)
{\mbox{\rm{tr}}} \left[ \Gamma_H \widetilde{S}_1(\hat{k}+w_1 \hat{p}) \Gamma_H
\widetilde{S}_2(\hat{k}-w_2 \hat{p}) \right] \,.
\label{selfenergy}
\end{eqnarray}

The requirement $Z_H=0$ implies that the physical state does not contain
the bare state and is appropriately described as a bound state. The interaction
leads to a dressed physical particle, i.e. its mass and wave function have to be
renormalized. The condition $Z_H=0$ also effectively excludes the constituent
degrees of freedom from the space of physical states. It thereby guarantees
the absence of double counting for the physical observable under consideration,
the constituents only exist in virtual states.

Therefore, the coupling constants $g_{H}$ of hadrons $H$ are strictly
fixed by the requirements of Eq.~(\ref{gren}) and do not constitute further free
parameters.

\section{One-photon radiative transitions of charmonium states}
\label{sec:transitions}

We consider the dominant one-photon radiative transitions of the charmonium
$S$- and $P$-wave states  $X= \{ J/\psi ({^{3}}\!S_{1}),  \chi_{c0}  ({^{3}}\!P_{0}),
\chi_{c1}  ({^{3}}\!P_{1}), h_{c}  ({^{1}}\!P_{1}), \chi_{c2}  ({^{3}}\!P_{2}) \}$.
The orbitally excited state $h_c(3525)$ is included in the considerations despite
the large uncertainty in its full decay width.

The latest data on the masses, full decay widths and branching fraction of the
dominant radiative decays of the charmonium S- and P-wave states reported
in \cite{PDG20} are given in Table~\ref{tab1}.

\begin{table}[H]
\begin{center}
\caption{The charmonium states ${^{2S+1}}\!{\it L}_{J}$  
and dominant radiative decay modes \cite{PDG20}.}
\label{tab1}
\begin{tabular}{|c|c|c|c|c|c|c|}
\hline
 State  & $J^{PC}$    & Interpolating current
              & Mass (MeV)  & Full width ($\Gamma$)  & Mode
              & Fraction ($\Gamma_{i} / \Gamma$)  \\
\hline
$\eta_c ({^{1}}\!S_{0}) $ & $0^{-+}$ & $i\bar q\, \gamma^5 q $
               & 2983.9$\pm$0.5 & 32.0$\pm$0.7 MeV & $\gamma \, \gamma$
               & (1.58$\pm$0.11) $\cdot 10^{-4}$ \\
$J/\psi ({^{3}}\!S_{1})$     &$1^{--}$  & $\bar q\, \gamma^\mu q $
               & 3096.9$\pm$0.0006 & 92.9$\pm$2.8 keV & $\gamma \, \eta_c$
               & (1.7$\pm$0.4)  $\cdot 10^{-2}$ \\
$\chi_{c0}  ({^{3}}\!P_{0})$ & $0^{++}$ & $\bar q\, I q $
               & 3414.71$\pm$0.30 & 10.8$\pm$0.6 MeV & $\gamma \, J/\psi$
               & (1.40$\pm$0.05)  $\cdot 10^{-2}$ \\
$\chi_{c1}  ({^{3}}\!P_{1})$ & $1^{++}$ & $\bar q\, \gamma^\mu\gamma^5\, q $
               & 3510.67$\pm$0.05 & 0.84$\pm$0.04 MeV &$\gamma \, J/\psi$
               &  (34.3$\pm$1.0)  $\cdot 10^{-2}$  \\
$h_{c}  ({^{1}}\!P_{1})$ & $1^{+-}$ & $i{\bar q}\!
                   \stackrel{\leftrightarrow}{\partial}_{\nu} \! \gamma^5  q $
               & 3525.38$\pm$0.11 & 0.7$\pm$0.4 MeV &$\gamma \, \eta_c$
               &  (51$\pm$6)  $\cdot 10^{-2}$  \\
$\chi_{c2}  ({^{3}}\!P_{2})$ & $2^{++}$ & ${\frac{i}{2}}\, {\bar q}\left(
                \gamma^\mu \!\! \stackrel{\leftrightarrow}{\partial}_{\nu} \!\! +
                \gamma^\nu \!\! \stackrel{\leftrightarrow}{\partial}_{\mu} \! \right) q $
              & 3556.17$\pm$0.07 & 1.97$\pm$0.09 MeV &$\gamma \, J/\psi$
              &  (19.0$\pm$0.5)  $\cdot 10^{-2}$  \\
\hline
\end{tabular}
\end{center}
\end{table}

The invariant matrix element for the one-photon radiative transition
$X_{1} \to \gamma X_{2}$ reads
\begin{eqnarray}
{\cal M}_{X_1 \to \gamma X_2}^\sigma = i(2\pi)^4\delta^{(4)}(p-q_1-q_2)\,
\varepsilon_{X_{1}}\,\varepsilon_{X_{2}}\,
\varepsilon_{\gamma\sigma} \, T_{X_1 \to \gamma X_2}^\sigma(q_1,q_2) \,,
\label{matrix}
\end{eqnarray}
where $\{p, q_1, q_2\}$ and $\{ \varepsilon_{X_{1}}\,,
\varepsilon_{X_{2}}\,,\varepsilon_\gamma\,\}$ are the
momenta and polarization vectors of the $X_{1}$, $X_{2}$ and the photon
($\gamma$), correspondingly.

\begin{figure}[H]
\begin{center}
\includegraphics[scale=0.25]{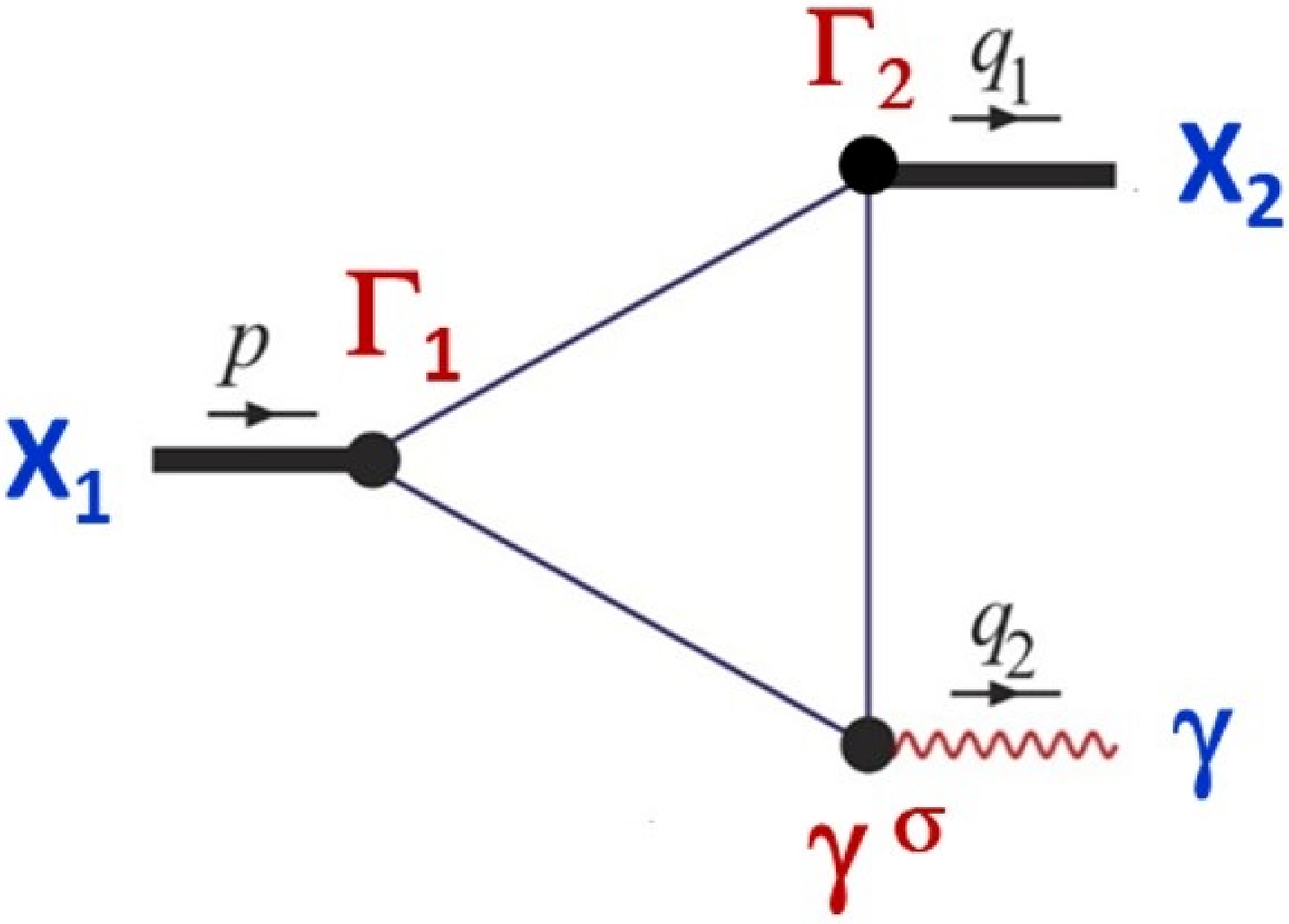}
\hspace*{2.5mm}
\includegraphics[scale=0.25]{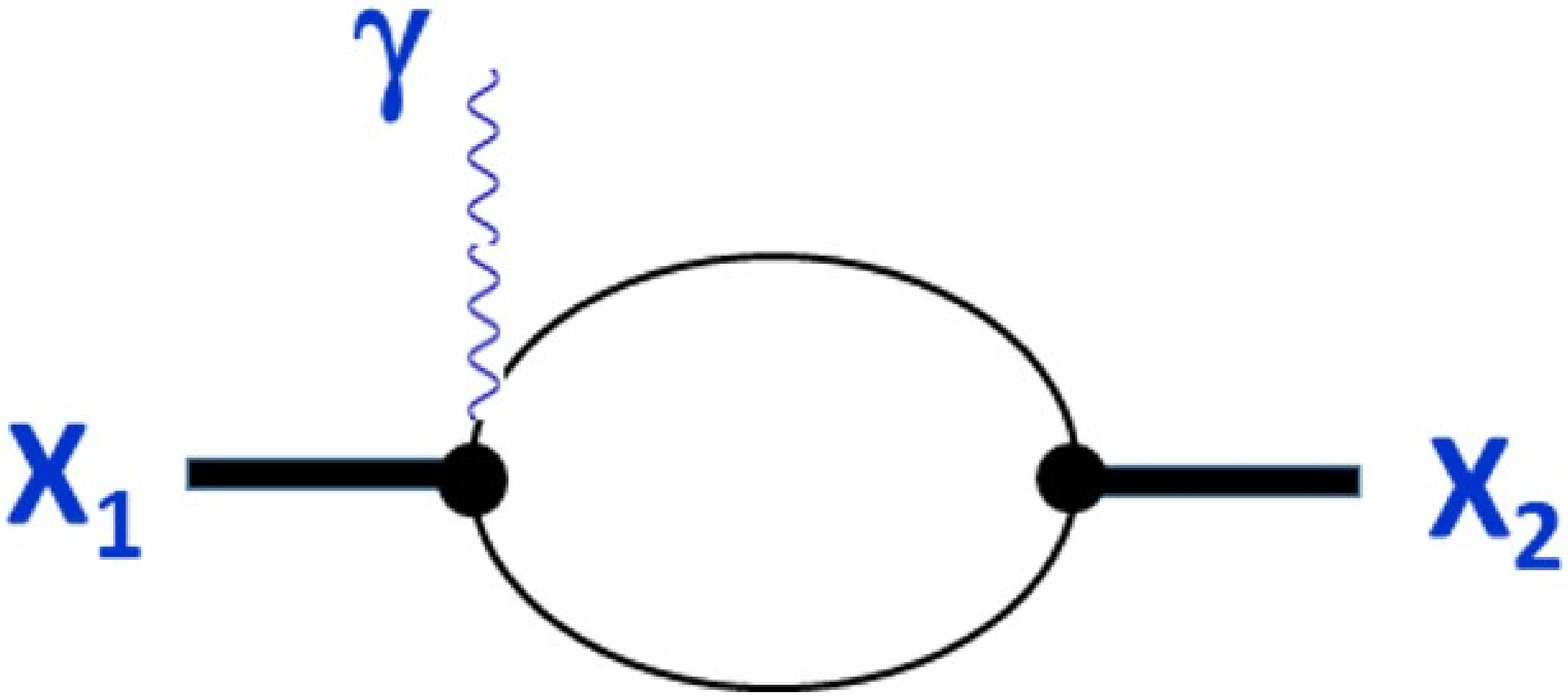}
\hspace*{2.55mm}
\includegraphics[scale=0.25]{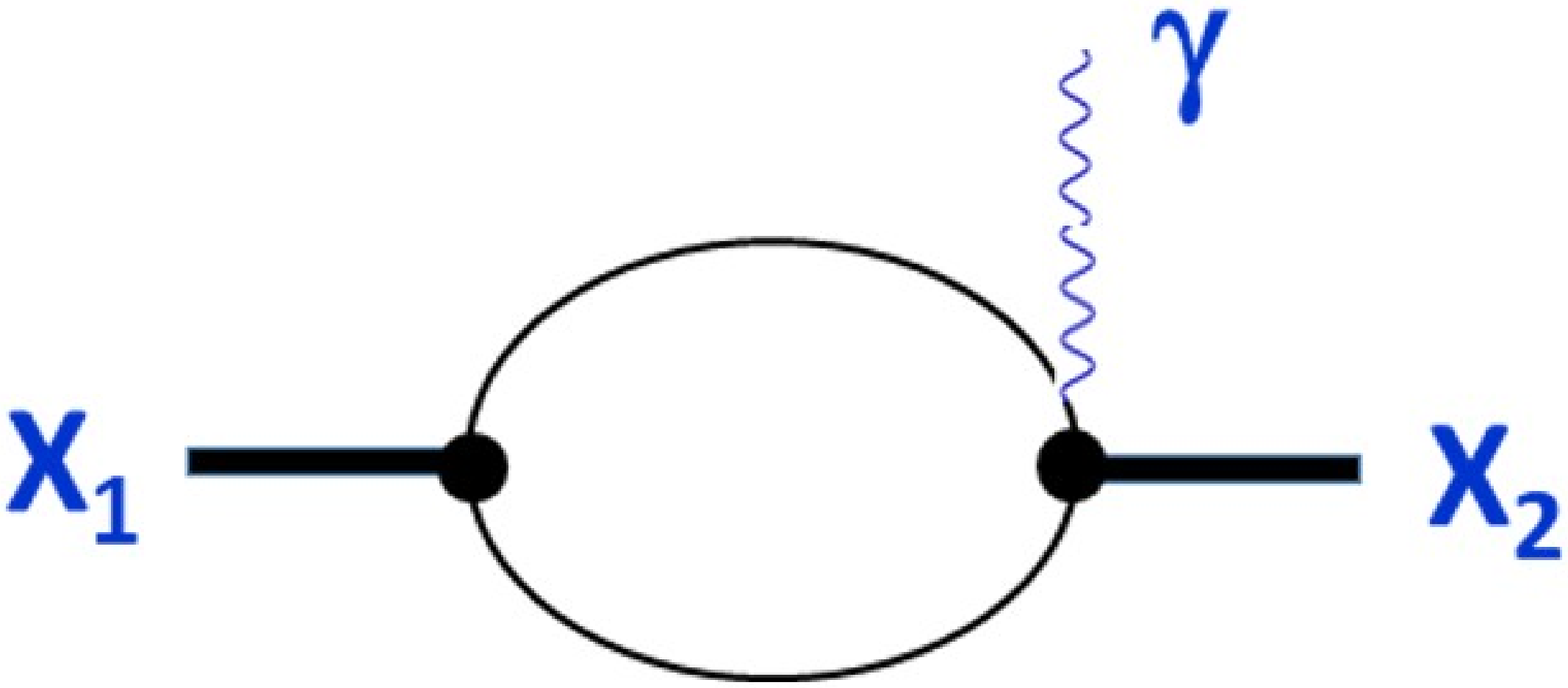}
\end{center}
\caption{
Feynman diagrams contributing in
leading order to the dominant one-photon radiative transitions
$X_{1}(p)\to\gamma(q_2)+X_{2}(q_1)$ of the charmonium states.
}
\label{fig2}
\end{figure}

In leading order, the transition amplitude $T_{X_1\to\gamma X_2}^\sigma(q_1,q_2)$
in Eq.~(\ref{matrix}) is described by the triangle and bubble Feynman
diagrams shown in Fig.~\ref{fig2}.

Note, each particular diagram in Fig.~\ref{fig2} is not gauge invariant by itself, 
but the total sum fulfills the gauge invariance requirement:   
\begin{eqnarray}
q_{2 \sigma} \cdot T^{\rm tot; \sigma}_{X_1 \to \gamma X_2}(q_1,q_2) = 0 \,.
\label{gaugeamp}
\end{eqnarray}
To simplify the calculation we split the contribution of each diagram 
$T^{\sigma}_{X_1 \to \gamma X_2}(q_1,q_2)$
into a part which is gauge invariant $T^{{\rm inv}; \sigma}_{X_1 \to \gamma X_2}(q_1,q_2)$
and one which is not $T^{{\rm rest}; \sigma}_{X_1 \to \gamma X_2}(q_1,q_2)$:  
\eq 
T^{\sigma}_{X_1 \to \gamma X_2}(q_1,q_2) = 
T^{{\rm inv}; \sigma}_{X_1 \to \gamma X_2}(q_1,q_2) \,+\, 
T^{{\rm rest}; \sigma}_{X_1 \to \gamma X_2}(q_1,q_2) \,. 
\en 
This separation can be achieved in the following manner. 
For the $\gamma$-matrix and the four momenta with photon Lorentz index we use the 
substitutions $\gamma^\sigma = \gamma^\sigma_\perp + q^\sigma\!\not\!\! q/q^2$ and 
$p^\sigma = p^\sigma_\perp + q^\sigma (p \cdot q)/q^2$ 
where $\gamma^\sigma_\perp = \gamma^\sigma - q^\sigma\!\not\!\! q/q^2$ and 
$p^\sigma_\perp = p^\sigma - q^\sigma (p \cdot q)/q^2$ 
obey the transversity conditions $\gamma^\sigma_\perp \cdot q_\sigma = 0$ and 
$p^\sigma_\perp \cdot q_\sigma = 0$. Expressions for diagrams containing only $\perp$-values 
are gauge invariant separately. It is easy to show that the remaining terms, which are not gauge 
invariant, cancel each other in total (see detailed discussion, e.g., in Ref.~\cite{Faessler:2003yf}). 
Therefore, it is enough to consider only the sum of the gauge-invariant contributions from 
all diagrams in Fig.~\ref{fig2}. The contributions to the full gauge-invariant amplitude given by 
the bubble-type diagrams are small and do not exceed the common errors ($\pm 10 \%$) of our
calculations within the CCQM.
Therefore, taking into account the uncertainty of the experimental data 
in Table~\ref{tab1}, we drop the bubble-type diagrams from the consideration
without loss in accuracy of our estimates.

The LO amplitude of the charmonium radiative (one-photon) transition described
by the gauge-invariant contribution of the triangle diagram in Fig.~2 in our approach reads 
\begin{eqnarray}
T_{X_1 \to \gamma X_2}^{{\rm inv}; \sigma}(q_1,q_2)
&\!\!=\!\!&
g_{X_1} g_{X_2}e_c e N_c \!\!\int\!\!\frac{d^4k}{(2\pi)^4i}
\widetilde{\Phi}_{X_1}\Big(\!\! -\! k^2\Big)
\widetilde{\Phi}_{X_2}\Big(\!\!-\!\!(k\!+\!\tfrac12 q_2)^2\Big)
\nonumber\\
&\!\!\!\cdot\!\!\!&
{\rm tr}\Big[\Gamma_2 S(\hat{k}+\tfrac12 \hat{p})
\Gamma_1 \widetilde{S}(\hat{k}-\tfrac12 \hat{p}) \gamma^\sigma_\perp 
\widetilde{S}(\hat{k}-\tfrac12 \hat{p}+\hat{q}_2)\Big] ,
\label{amp0}
\end{eqnarray}
where $e_c\!=\! 2/3$, $e$ is the electron electric charge;
$ \Gamma_{1}\!=\!\{\gamma^{\mu}, I, \gamma^{\mu}\gamma_5,\,
\stackrel{\leftrightarrow}{\partial}_{\nu}\!\!\gamma^5\,,
i ( \gamma^\mu \!\!\stackrel{\leftrightarrow}{\partial}_{\nu}
+\gamma^\nu \!\!\stackrel{\leftrightarrow}{\partial}_{\mu} )/2 \}$
for $\{J/\psi,\chi_{c0},\chi_{c1},h_c,\chi_{c2}\}$
and $ \Gamma_{2}\!=\!\{ i \gamma^5, \gamma^{\mu} \}$
for $\{\eta_c,J/\psi\}$, correspondingly.

As mentioned above, the renormalized couplings $g_{X_1}$ and $g_{X_2}$
are strictly determined  by the self-energy (mass) function of the
corresponding charmonia in Eq.~(\ref{gren}).

By substituting the Gaussian-type vertices functions and the charm quark
propagators in the Fock-Schwinger representation we rewrite the
transition amplitude as follows:
\begin{eqnarray}
T_{X_1 \to \gamma X_2}^{{\rm inv}; \sigma}(q_1,q_2)
&\!\!\!=\!\!\!& g_{X_1} g_{X_2}e_c e N_c
\!\!\int\!\!\!\!\intop_{0}^{1}\!\!\!\!\int\!\! d\alpha_1\,d\alpha_2\,d\alpha_3
\nonumber\\
&\!\!\cdot\!\!&
\!\!\int\!\!\frac{d^4k}{(2\pi)^4i}
\exp\left\{ \! k^2 (\alpha_1 \!+ \alpha_2 \!+ \alpha_3 \!+ s_1 \!+ s_2)
+ 2k^\nu R^\nu + R_0 \right\}
\nonumber\\
&\!\!\!\cdot\!\!\!&
{\rm tr}\Big[\Gamma_2 (m_c+\hat{k}+\tfrac12 \hat{p})
\, \Gamma_1 (m_c+\hat{k}-\tfrac12 \hat{p}) \gamma^\sigma_\perp 
(m_c+\hat{k}-\tfrac12 \hat{p}+\hat{q}_2)\Big] \,. 
\label{amp}
\end{eqnarray}
Here, $R^\nu$ and $R_0$ are linear combinations of the external momenta
$\{ p,q_1,q_2 \}$ which also depend on the charmonium size parameters $\{ s_1,s_2\}$
as well as the integration variables $\{\alpha_1, \alpha_2, \alpha_3\}$.

Note, we use the same Gaussian-type vertex function [see Eq.~(\ref{vertex})]
both for the ground-state and orbitally excited charmonia.

We take the trace for given $\Gamma_1$ and $\Gamma_2$ and
perform an explicit $k$-integration in Eq.~(\ref{amp}) while turning
the set of Fock-Schwinger parameters into a simplex (see Eq.~(\ref{integ1})). Finally
we obtain the LO gauge-invariant transition amplitude of the dominant radiative
decay of a charmonium state as follows:
\begin{eqnarray}
T_{X_1 \to \gamma X_2}^{{\rm inv}}(q_1,q_2)
\!\!&=&\!\!
\frac{g_{X_1}\, g_{X_2}\,e_c\, e\, N_c}{(2\pi)^2}\,
\intop_{0}^{1/\lambda^2}  \!\!\! dt \frac{t^2}{(s+t)^2}
\int\!\!\!\intop_{0}^{1}\!\!\!\int\!\!\! d\alpha_1\,d\alpha_2\,d\alpha_3
\,\delta(1-\alpha_1-\alpha_2-\alpha_3)
\nonumber\\
\!\!&\cdot&\!\!
\!\!f_{\Gamma_1,\Gamma_2}(p,q_1,q_2,m_c,s,t,\alpha_1,\alpha_2,\alpha_3)
\cdot \exp\Big( \! - \! t\,z_0 \!+\! \frac{t\,s}{s+t}\,z_1 \!+\! \frac{s^2}{s+t}\,z_2 \!\Big),
\label{amp2}
\end{eqnarray}
where
\begin{eqnarray*}
z_0&=& m_c^2  - p^2\,\alpha_1\alpha_2 - q_1^2\,\alpha_2\alpha_3\,,
\nonumber\\
z_1&=& p^2 (\alpha_1-\bar{s}_{1}/2)(1/2-\alpha_2)
    + q_1^2 (\alpha_1-1/2)(\alpha_1+\alpha_2-\bar{s}_{1} - \bar{s}_{2} /2)\,,
\nonumber\\
z_2&=& q_2^2 \bar{s}_{1} \bar{s}_{2} /4\,,
\nonumber\\
s_1&=&1/\Lambda^2_{X_1} \,,  \quad s_2=1/\Lambda^2_{X_2}\,, \quad
s=s_1+s_2\,, \quad  \bar{s}_1=s_1/s\,, \quad \bar{s}_2=s_2/s .
\end{eqnarray*}
$\Lambda_{X_1}$ and $\Lambda_{X_2}$ are the size parameters for the
in and outgoing charmonia, while $p^2=M_{X_1}^2$, $q_1^2=M_{X_2}^2$
have on-mass-shell values.
The limit $q_2^2 = 0$ corresponds to  an outgoing real photon in the radiative decay.

Note, the kernel integrand $f_{\Gamma_{1},\Gamma_{2}}(\ldots)$  in
Eq.~(\ref{amp2}) is formed by the trace operation in Eq.~(\ref{amp}),
and defines the main structure of the transition amplitude.
Below we apply the CCQM to the charmonium S- and P-wave states where explicit expressions
for the transition amplitudes are presented.

\subsection*{Transition $J/\psi({}^3\!S_1) \to \gamma \eta_c({}^1\!S_0)$}

First we consider a typical electromagnetic M1 transition between ground
states, from the vector $J/\psi$ ($\Gamma_1=\gamma^\rho$) to the pseudoscalar
$\eta_c$ ($\Gamma_2=i\gamma^5$) by radiating a photon
($\Gamma_\gamma=\gamma^{\sigma}$).

For this case we calculate in Eq.~(\ref{amp2}):
\eq
f_{\Gamma_1,\Gamma_2} = m_c \, \epsilon^{q_1 q_2 \rho \sigma} \,,
\qquad \epsilon^{q_1 q_2 \rho \sigma}
\equiv \epsilon^{\mu \nu \rho \sigma} q_1^\mu q_2^\nu \,,
\en
where $\epsilon^{\mu \nu \rho \sigma}$ is the four-dimensional antisymmetric
Levi-Civita tensor.

Accordingly, the gauge invariant transition amplitude in Eq.~(\ref{amp2})
takes the form:
\eq
T^{{\rm inv}; \rho\sigma}_{J/\psi \to \gamma \eta_c}
= g_{J/\psi} \, g_{\eta_c} \, C(p^2,q_1^2,q_2^2)\,\epsilon^{q_1 q_2 \rho \sigma} \,,
\label{ampJpsi}
\en
where the form factor reads:
\eq
C = \frac{e_c e N_c m_c}{(2\pi)^2}
\intop_{0}^{1/\lambda^2} \!\!\!  \frac{dt \, t^2}{(s+t)^2}
\!\!\intop_{0}^{1}\!\!\!\!\int\!\! d\alpha_1 d\alpha_2  \,
\exp\left\{ -  t\,z_0 \!+\! \frac{t\,s}{s+t}\,z_1 \!+\! \frac{s^2}{s+t}\,z_2 \!
\right\} \,,
\label{formJpsi}
\en
with $s_1=1/\Lambda_{J/\psi}^2$, $s_2=1/\Lambda_{\eta_c}^2$,
$p^2=M^2_{J/\psi}$, $q_1^2=M^2_{\eta_c}$, $q_2^2=0$ and
$\alpha_3=1-\alpha_1-\alpha_2$.

The renormalized couplings $g_{J/\psi}$ and  $g_{\eta_c}$ are dimensionless
and defined according to Eq.~(\ref{gren}) by using the following diagonal
parts of the corresponding polarization kernels:
\eq
g_{J/\psi} \!\!&=&\!\!
\left\{
\frac{N_c}{3}\Big( g_{\mu\nu} - \frac{p_\mu \, p_\nu}{p^2}\Big)
\!\! \int\!\!\!\frac{d^4k}{(2\,\pi)^4\,i}\, \widetilde\Phi_{J/\psi}^2(-k^2)\,
{\rm tr}\left[\gamma_\mu\,\widetilde S(\hat{k}-\frac{\hat{p}}{2})\,\gamma_\nu\,
\widetilde S(\hat{k}+\frac{\hat{p}}{2})\right]
\right\}^{-1/2} \!\!\!\!\!, \,\,\, p^2=M^2_{J/\psi},
\nonumber\\
g_{\eta_c} \!\!&=&\!\!
\left\{
N_c \!\! \int\!\!\!\frac{d^4k}{(2\,\pi)^4\,i}\, \widetilde\Phi_{\eta_c}^2(-k^2)\,
{\rm tr}\left[\gamma^5\,\widetilde S(\hat{k}-\frac{\hat{p}}{2})\,\gamma^5\,
\widetilde S(\hat{k}+\frac{\hat{p}}{2})\right]
\right\}^{-1/2} \!\!\!\!\!, \quad p^2=M^2_{\eta_c} \,.
\label{grenJpsi}
\en

Finally, we calculate the fractional one-photon radiative decay width of the
vector ground state $J/\psi$ as follows:
\begin{eqnarray}
\Gamma(J/\psi \to \gamma \eta_c) = \frac{\alpha}{24} \,
g_{J/\psi}^2 \, g_{\eta_c}^2 \, M_{J/\psi}^3 \,
\left( 1- \frac{M_{\eta_c}^2}{M_{J/\psi}^2} \right)^3
\cdot \left[ C(M^2_{J/\psi},M^2_{\eta_c},0)\right]^2 \,,
\label{widthJpsi}
\end{eqnarray}
where $\alpha=e^2/4\pi =1/137.036$ is the fine-structure constant. 

\subsection*{Transition $ \chi_{c0}({^{3}}\!P_{0}) \to \gamma J/\psi({}^3\!S_1)$}

The most general form of the one-photon radiative transition amplitude of the
orbitally excited (scalar) charmonium into the vector ground-state reads:
\eq
T^{{\rm inv}; \rho\sigma}_{\chi_{c0} \to \gamma J/\psi}(q_1,q_2)
= a\,g^{\rho\sigma} + b\,q_1^{\rho}\,q_1^{\sigma}
 + c\,q_1^{\rho}\,q_2^{\sigma} + d\,q_2^{\rho}\,q_1^{\sigma}
 + e\,q_2^{\rho}\,q_2^{\sigma}
\en
with five seemingly independent terms (see, e.g.~\cite{game09}).

By taking into account the Lorentz conditions
($q_1^{\rho}\cdot \varepsilon_{J/\psi; \rho} = 0$,~$q_2^{\sigma}\cdot
\varepsilon_{\gamma; \sigma} = 0$) and applying the gauge invariance condition
($q_{2\sigma} \cdot T^{{\rm inv}; \rho\sigma}_{\chi_{c0}\to\gamma J/\psi}=0$)
one finds that the terms with coefficients $b$, $c$, and $e$ do not contribute
while two remaining coefficients are related: $a = - d \, (q_1\cdot q_2)$.

Therefore, the number of independent terms is reduced to one and the
gauge-invariant transition amplitude reads
\eq
T^{{\rm inv}; \rho\sigma}_{\chi_{c0}\to\gamma J/\psi}(q_1,q_2)
= g_{\chi_{c0}} \, g_{J/\psi} \, d(p^2,q_1^2,q_2^2)
\cdot \big(   q_1^\sigma q_2^\rho - g_{\rho\sigma}(q_1\!\cdot\! q_2)\big)
\label{ampXc0}
\en
resulting in
\eq
| \varepsilon_{J/\Psi; \rho}\, \varepsilon_{\gamma; \sigma}\,
T^{{\rm inv}; \rho\sigma}_{\chi_{c0}\to\gamma J/\psi}(q_1,q_2) |^2
= \frac{1}{2} M^4_{\chi_{c0}}
\big( 1 - M^2_{\Psi}/M^2_{\chi_{c0}}  \big)^2 \, g_{\chi_{c0}}^2 \,  g_{J/\psi}^2
\,  d^2 \,.
\label{matrixXc0}
\en

For $\Gamma_1 = I$ and $\Gamma_2 = \gamma^{\rho}$ we calculate
in Eq.~(\ref{amp2}):
\eq
f_{\Gamma_1,\Gamma_2} = m_c \, \left( \frac{1}{2}
+ \frac{t\alpha_3+s_2/2}{s+t} \right) \,
\cdot \big(   q_1^\sigma q_2^\rho - g_{\rho\sigma}(q_1\!\cdot\! q_2)\big)
\en
and accordingly, the form factor $d$ reads:
\eq
d = \frac{e_c e N_c m_c}{(2\pi)^2}
\intop_{0}^{1/\lambda^2} \!\!\! dt \frac{t^2}{(s+t)^2}
\!\!\intop_{0}^{1}\!\!\!\!\int\!\! d\alpha_1 d\alpha_2
\left( \frac{1}{2} + \frac{t\alpha_3+s_2/2}{s+t} \right)
e^{-t\,z_0 + \frac{t\,s}{s+t}\,z_1} \,,
\label{formXc0}
\en
where $s_1=1/\Lambda_{\chi_{c0}}^2$, $s_2=1/\Lambda_{J/\psi}^2$,
$\alpha_3=1-\alpha_1-\alpha_2$,
$p^2=M^2_{\chi_{c0}}$, $q_1^2=M^2_{J/\psi}$ and $q_2^2=0$.

The renormalized coupling  $g_{\chi_{c0}}$ is dimensionless and defined
according to Eq.~(\ref{gren}) by using the mass function as follows:
\eq
g_{\chi_{c0}}\!\!&=&\!\!
\left\{
N_c \int\!\!\!\frac{d^4k}{(2\,\pi)^4\,i}\, \widetilde\Phi_{\chi_{c0}}^2(-k^2)\,
{\rm tr}\left[ \widetilde S(k-\frac{p}{2})\, \widetilde S(k+\frac{p}{2})\right]
\right\}^{-1/2} \,.
\label{grenXc0}
\en

The partial width of the decay $\chi_{c0} \to \gamma J/\psi$ is obtained
as follows:
\eq
\Gamma(\chi_{c0} \to \gamma J/\psi) =
\frac{\alpha}{24} g_{\chi_{c0}}^2 \,  g_{J/\psi}^2 \, M_{\chi_{c0}}^3
\left( 1- \frac{M_{J/\psi}^2}{M_{\chi_{c0}}^2} \right)^3
\cdot \left[ d(g_{\chi_{c0}},g_{J/\psi},M^2_{\chi_{c0}},M^2_{J/\psi},0)\right]^2 \,.
\label{widthXc0}
\en

\subsection*{Transition $ \chi_{c1}({^{3}}\!P_{1}) \to \gamma J/\psi({}^3\!S_1)$}

The matrix element for the transition $\chi_{c1} \to \gamma J/\psi$ of the
excited axial-vector charmonium ($\Gamma_1=\gamma^\mu\gamma^5$)
into the vector ground state ($\Gamma_2=\gamma^\rho$)  reads
\eq
{\cal M}_{\chi_{c1}\to\gamma J/\psi} \sim \varepsilon_{\chi_{c1; \mu}}(p)
\cdot \varepsilon_{J/\psi; \rho}(q_1)\cdot \varepsilon_{\gamma; \sigma}(q_2)
\cdot T^{{\rm inv}; \mu\rho\sigma}_{\chi_{c1}\to\gamma J/\psi} \,,
\en
where the polarization vectors $\varepsilon_{\chi_{c1}}$, $\varepsilon_{J/\psi}$
and $\varepsilon_{\gamma}$ satisfy transversality, completeness and
orthonormality conditions (see, e.g., Ref.~\cite{bran10ivan17}).

By taking into account the Lorentz conditions
\eq
p_\mu       \, \varepsilon_{\chi_{c1}}^\mu=0, \qquad 
q_{1\rho}   \, \varepsilon_{J/\psi}^\rho  =0,  \qquad 
q_{2\sigma} \, \varepsilon_\gamma^\sigma  =0
\en
and applying  the gauge invariance requirement
\eq
q_{2\sigma} \, T^{{\rm inv}; \mu\rho\sigma}_{\chi_{c1}\to\gamma J/\psi}(q_1,q_2) = 0
\en
we write down the gauge-invariant transition amplitude with four seemingly
independent Lorentz structures as follows:
\eq
T^{{\rm inv}; \mu\rho\sigma}_{\chi_{c1}\to\gamma J/\psi}(q_1,q_2)
&=& g_{\chi_{c1}} \,  g_{J/\psi} \, \left[
\epsilon^{q_2\mu\sigma\rho} (q_1\cdot q_2)\,W_1
         +\epsilon^{q_1q_2\sigma\rho} q_1^\mu\,W_2 \right.
\nonumber\\
&+& \left. \epsilon^{q_1q_2\mu\rho} q_2^\sigma\,W_3
       +\epsilon^{q_1q_2\mu\sigma} q_1^\rho\,W_4
        -\epsilon^{q_1\mu\sigma\rho} (q_1\cdot q_2)\,W_4 \right] \,.
\label{ampXc1a}
\en
In accordance with this pattern, we calculate the form factors $W_1$,  $W_2$, $W_3$
and  $W_4$ from  Eq.~(\ref{amp2}) by taking into account definitions
$s_1=1/\Lambda_{\chi_{c1}}^2$, $s_2=1/\Lambda_{J/\psi}^2$,
$\alpha_3=1-\alpha_1-\alpha_2$ and the on-mass-shell conditions
$p^2=M^2_{\chi_{c1}}$, $q_1^2=M^2_{J/\psi}$, $q_2^2=0$.

These form factors $W_i$ are proportional to the dimensionless renormalized
coupling  $g_{\chi_{c1}}$ which is defined [see Eq.~(\ref{gren})] by using the
diagonalized polarization kernel:
\eq
g_{\chi_{c1}}\!\!&=&\!\!
\left\{
\frac{N_c}{3}\Big( g_{\mu\nu} - \frac{p_\mu \, p_\nu}{p^2}\Big)
\int\!\!\!\frac{d^4k}{(2\,\pi)^4\,i}\, \widetilde\Phi_{\chi_{c1}}^2(-k^2)\,
{\rm tr}\left[\gamma^\mu \gamma^5  \,\widetilde S(k-\frac{p}{2})\,
\gamma^\nu \gamma^5\, \widetilde S(k+\frac{p}{2})\right]
\right\}^{-1/2} \!\!\!\!\! \,.
\label{grenXc1}
\en

By using specific symmetry relations between tensors
$\epsilon^{\mu\nu\rho\sigma}$ described earlier in \cite{dubn10} we can
reduce the number of independent Lorentz structures into two as follows:
\eq
T^{{\rm inv}; \mu\rho\sigma}_{\chi_{c1}\to\gamma J/\psi} =
g_{\chi_{c1}} \,  g_{J/\psi} \, \left[
\epsilon^{q_1q_2\mu\rho} \, q_2^\sigma
\left(W_{1}\!+\!W_{3}\!-\!\frac{q_1^2}{(q_{1}\, q_{2})}W_{4}\right)
\!+\!
\epsilon^{q_1q_2\sigma\rho} \, q_1^\mu
\left(W_{1}\!+\!W_{2}\!-\! W_4\!-\!\frac{q_1^2}{(q_{1}\, q_{2})} W_{4}\right)
\right]\,.
\label{ampXc1}
\en

We further introduce the following helicity amplitudes (see, \cite{dubn10})
\eq
H_L &=& i g_{\chi_{c1}} \,  g_{J/\psi} \,\frac{M_{\chi_{c1}}^{2}}{M_{J/\psi}}|
     \vec q_2|^{2}
     \Big[ W_1 + W_3-\frac{M^2_{J/\psi}}{M_{\chi_{c1}}|\vec q_2|}W_4\Big]\,,
     \qquad  |\vec q_2|=\frac{M_{\chi_{c1}}^2-M^2_{J/\psi}}{2M_{\chi_{c1}}}\,,
\nonumber\\
H_T &=& -i g_{\chi_{c1}} \,  g_{J/\psi} \,M_{\chi_{c1}}|\vec q_2|^{2}
      \Big[W_1+W_2
        -\Big(1+\frac{M^2_{J/\psi}}{M_{\chi_{c1}}|\vec q_2|}\Big)\,W_4\Big] \,.
\label{helicity}
\en

Then, the partial decay width of the axial-vector charmonium
$\chi_{c1}({^{3}}\!P_{1})$ reads
\eq
\Gamma(\chi_{c1}\to \gamma J/\Psi) =
\frac{g_{\chi_{c1}}^2 \,  g_{J/\psi}^2 \,}{12\pi} \frac{|\vec q_2|}{M_{\chi_{c1}}^2}
\Big( |H_L|^2 +|H_{T}|^{2}\Big) \,.
\label{widthXc1}
\en

\subsection*{Transition $ h_c({^{1}}\!P_{1}) \to \gamma \eta_c ({^{1}}\!S_{0})  $}

As mentioned above, we also consider the one-photon radiative
transition of the orbitally excited charmonium state $h_c(3525)$ despite the
large uncertainty in the full decay width (see Table I).

By substituting the matrices $\Gamma_1 = k^\mu  \gamma^5$ in Eq.~(\ref{amp})
and $\Gamma_2 = \gamma^{\rho}$ we obtain the gauge invariant amplitude of
the transition  $h_c \to \gamma \eta_c$  as follows:
\eq
T^{{\rm inv}; \rho\sigma}_{h_{c}\to\gamma\eta_c}(q_1,q_2)
=  g_{h_c} \, g_{\eta_c} \, {\emph h} (p^2,q_1^2,q_2^2) \cdot
( q_2^\rho q_1^\sigma - g^{\rho\sigma}(q_1\cdot q_2)\big)\,,
\label{ampHc}
\en
where the form factor ${\emph h}$ is calculated by using Eq.~(\ref{amp}) in
a similar way as described in the previous subsections.

In contrast to $g_{\eta_c}$ the renormalized coupling  $g_{h_c}$ is dimensional
($\sim$ GeV$^{-1}$) and defined [see Eq.~(\ref{gren})] by using the
diagonalized polarization kernel:
\eq
g_{h_c}\!\!&=&\!\!
\left\{
\frac{N_c}{3}\Big( g_{\mu\nu} - \frac{p_\mu \, p_\nu}{p^2}\Big)
 \int\!\!\!\frac{d^4k}{(2\,\pi)^4\,i}\, \widetilde\Phi_{h_c}^2(-k^2)\,
{\rm tr}\left[k^\mu \gamma^5  \,\widetilde S(k-\frac{p}{2})\,
k^\nu \gamma^5\, \widetilde S(k+\frac{p}{2})\right]
\right\}^{-1/2}\!\!\!\! .
\label{grenHc}
\en

For the square of the invariant matrix element we have
\eq
| {\cal M}_{h_{c}\to\gamma\eta_c} |^2
\sim
|\varepsilon_{h_c; \rho}\, \varepsilon_{\gamma; \sigma}\,
T^{{\rm inv}; \rho\sigma}_{h_{c}\to\gamma\eta_c}(q_1,q_2) |^2
 = \frac{1}{2} g_{h_c}^2 \, g_{\eta_c}^2 \,
 M^4_{h_c} \big( 1 - M^2_{\eta_c}/M^2_{h_c}  \big)^2
 \cdot |{\emph h}  (p^2,q_1^2,q_2^2)|^2
\label{matrixHc}
\en
with $s_1=1/\Lambda_{h_{c}}^2$, $s_2=1/\Lambda_{J/\psi}^2$,
$p^2=M^2_{h_{c}}$, $q_1^2=M^2_{J/\psi}$ and $q_2^2=0$.

The fractional decay width of the excited charmonium state
$h_{c}({^{1}}\!P_{1})$ reads
\eq
\Gamma(h_{c} \to \gamma \eta_c) =
\frac{\alpha\,  g_{h_c}^2 \, g_{\eta_c}^2 \,}{24\,(1+2S)} M_{h_{c}}^3
\left( 1- \frac{M_{\eta_c}^2}{M_{h_{c}}^2} \right)^3
 \cdot |{\emph h}  (M_{h_{c}}^2,M_{\eta_c}^2,0)|^2 \,,
\label{widthHc}
\en
with the spin value $S=1$.

\subsection*{Transition $ \chi_{c2}({^{3}}\!P_{2}) \to \gamma J/\psi({}^3\!S_1)$}

The polarization vector $\varepsilon_{\mu\nu}^{(\lambda)}$ (below we denote
as $\varepsilon^{\mu\nu}_{\chi_{c2}}$) of the tensor meson $\chi_{c2}$ satisfies
the symmetry, transversality, tracelessness, orthonormality and completeness
conditions:
\eq
\varepsilon^{(\lambda)}_{\mu\nu}(p) = \varepsilon^{(\lambda)}_{\nu\mu}(p),
\qquad
\varepsilon^{(\lambda)}_{\mu\nu}(p)\, p^\mu = 0,
\qquad
\varepsilon^{(\lambda)}_{\mu\nu}(p)\, g^{\mu\nu} = 0,
\qquad
\varepsilon^{\dagger\,(\lambda)}_{\mu\nu} \varepsilon^{(\lambda')\,\mu\nu}
=\delta_{\lambda \lambda'},
\nonumber\\
\sum\limits_{\lambda=0,\pm 1, \pm 2} \!\!\!
\varepsilon^{(\lambda)}_{\mu\nu}
\varepsilon^{\dagger\,(\lambda)}_{\alpha\beta} = \frac{1}{2}
\left(\bar{g}_{\mu\alpha}\,\bar{g}_{\nu\beta}
+ \bar{g}_{\mu\beta}\,\bar{g}_{\nu\alpha}\right)
- \frac{1}{3}\,\bar{g}_{\mu\nu}\,\bar{g}_{\alpha\beta}
\equiv \widetilde{g}_{\mu\nu\alpha\beta},   \quad
\bar{g}_{\mu\nu} \equiv -g_{\mu\nu}+\frac{p_\mu\,p_\nu}{p^2} \,.
\label{polarXc2}
\en

For the gauge invariant amplitude of the transition $\chi_{c2} \to \gamma J/\psi$
we substitute the matrices $\Gamma_1 = \gamma^\mu k^\nu \!+\!\gamma^\nu k^\mu$
and $\Gamma_2 = \gamma^{\rho}$ into Eq.~(\ref{amp}) and obtain
\eq
T^{{\rm inv}; \mu\nu\rho\sigma}_{\chi_{c2}\to\gamma J/\psi} (q_1,q_2) &\!\!=\!\!&
g_{\chi_{c2}} \,  g_{J/\psi} \, \left\{
A\cdot \Big( g^{\mu\rho} \Big[ g^{\sigma\nu} (q_1\cdot q_2) - q_1^{\sigma}\,q_2^{\nu} \Big]
           + g^{\nu\rho} \Big[ g^{\sigma\mu} (q_1\cdot q_2) - q_1^{\sigma}\,q_2^{\mu} \Big] \Big)
           \right.
\nonumber\\
&\!\!+\!\!& \left.
B\cdot \Big( g^{\sigma\rho} \Big[ q_1^{\mu}\,q_2^{\nu} + q_1^{\nu}\,q_2^{\mu} \Big]
         - g^{\mu\sigma} q_1^{\nu}\,q_2^{\rho} - g^{\nu\sigma} q_1^{\mu}\,q_2^{\rho} \Big)
         \right\} \,,
\label{ampXc2}
\en
where the two independent form factors $A(g_{\chi_{c2}},g_{J/\psi},p^2,q_1^2,q_2^2)$
and $B(g_{\chi_{c2}},g_{J/\psi},p^2,q_1^2,q_2^2)$ are determined
by Eq.~(\ref{amp2}) with
$s_1=1/\Lambda_{\chi_{c2}}^2$, $s_2=1/\Lambda_{J/\psi}^2$,
$p^2=M^2_{\chi_{c2}}$, $q_1^2=M^2_{J/\psi}$ and $q_2^2=0$.

The renormalized coupling of the tensor meson is defined [see Eq.~(\ref{gren})]
as follows:
\eq
g_{\chi_{c2}} \!\!&=&\!\!
\left\{ \frac{\widetilde{g}_{\mu\nu\rho\sigma}}{5}\,
\!\!\!\int\!\!\!\!\frac{d^4k}{(2\,\pi)^4\,i}\, \widetilde\Phi_{\chi_{c2}}^2(-k^2)\, {\rm tr}
\left[
\left(\gamma^\mu k^\nu \!+\!\gamma^\nu k^\mu\right) \widetilde{S}(k-\frac{p}{2})
\left(\gamma^\rho k^\sigma \!+\!\gamma^\sigma k^\rho\right)
\widetilde{S}(k+\frac{p}{2}) \right] \right\}^{-1/2} \,
\label{grenXc2}
\en
with dimension of ($\sim$ GeV$^{-1}$).

For the square of the invariant matrix element for the radiative transition
of the tensor charmonium $\chi_{c2} \to \gamma J/\Psi$ we obtain
\eq
| {\cal M}_{\chi_{c2}\to\gamma J/\psi} |^2
\sim | \varepsilon_{\chi_{c2}; \mu\nu} \varepsilon_{J/\Psi; \rho}\,
\varepsilon_{\gamma; \sigma}\,
T^{{\rm inv}; \mu\nu\rho\sigma}_{\chi_{c2}\to\gamma J/\psi} (q_1,q_2) |^2
\!=\! g_{\chi_{c2}}^2 \,  g_{J/\psi}^2 \, M_{\chi_{c2}}^4
\big( C_{A} \!\cdot\! A^2 \!+\! C_{AB}\!\cdot\! A \!\cdot\! B
\!+\! C_{B}\!\cdot\! B^2 \big) \,,
\label{matrixXc2}
\en
where the numerical constants $C_{A}$, $C_{AB}$ and $C_{B}$ are defined
through the meson masses as follows:
\eq
C_{A} &=& \frac{1}{\xi}  \left( \frac{1}{4} + \frac{7}{3}\xi - \frac{31}{6}\xi^2
                  + \frac{7}{3}\xi^3 + \frac{1}{4}\xi^4 \right) = 0.1957,
                  \qquad \xi = M_{J/\psi}^2 / M_{\chi_{c2}}^2 = 0.7584,
\nonumber\\
C_{AB} &=& - \frac{1}{\xi} \left( \frac{1}{2} - \frac{1}{3}\xi - 2\xi^2 +3\xi^3
                  - \frac{7}{6}\xi^4 \right) = - 0.02576,
\\
C_{B} &=&  \frac{1}{\xi} \left( \frac{1}{4} - \frac{2}{3}\xi + \frac{1}{6}\xi^2 + \xi^3
                  - \frac{13}{12}\xi^4 + \frac{1}{3}\xi^5 \right) = 0.00226 \,.
\nonumber
\label{formXc2}
\en

The fractional decay width of the one-photon radiative transition of the
tensor charmonium excitation $\chi_{c2}$ reads:
\eq
\Gamma(\chi_{c2} \to \gamma J/\psi)
= \frac{\alpha \, g_{\chi_{c2}}^2 \,  g_{J/\psi}^2 \,}{4 (1\!+\!2S)}\,
M_{\chi_{c2}}^3  \left( 1- \frac{M_{J/\psi}^2}{M_{\chi_{c2}}^2} \right)
\cdot \big( C_{A} \!\cdot\! A^2 \!+\! C_{AB}\!\cdot\! A \!\cdot\! B
\!+\! C_{B}\!\cdot\! B^2 \big)
\label{widthXc2}
\en
with spin value $S=2$.

\section{Numerical results}
\label{sec:numerical}

According to the CCQM the hadronic field is coupled to a nonlocal quark
current by the interaction Lagrangian described in Eq. (\ref{Lagrangian}),
where the nonlocal vertex function in Eq. (\ref{vertex}) characterizes the
quark distribution inside the hadron. The vertex function
$\widetilde{\Phi}_H\left(-p^2\right)$ is unique for the given hadron and each
hadron has its own adjustable parameter $\Lambda_H$, which can be related
to the hadron 'size'. The fixed values of the ``size'' parameters of various
hadrons differ in the CCQM but for most cases a pattern can be traced
- the heavier a hadron, the larger its ``size''.

On the other hand, the charmonium members under consideration have
the same quark content ($c\bar{c}$) and possess physical masses in a relative
narrow interval $\sim 3 \div 3.5$ GeV. Therefore, for this specific case we
use the ansatz that the charmonium 'size' is proportional to its physical
mass, i.e.,  $\Lambda_X = \varrho\! \cdot \!\! M_X$ with a constant
$\varrho >0$.

Subsequently, we introduce one common and adjustable ``slope'' parameter:
\eq
\varrho \equiv \Lambda_X/M_X \,,
\en
instead of six separate 'size' parameters $\Lambda_{X} = \{ \Lambda_{\eta_{c}},
\Lambda_{J/\psi}, \Lambda_{\chi_{c0}}, \Lambda_{\chi_{c1}}, \Lambda_{h_{c}},
\Lambda_{\chi_{c2}} \}$ to describe the quark distribution inside the charmonia.

We further use the charmonium vertex function defined as:
\eq
\widetilde{\Phi}_X\left(-p^2\right)
= \exp\left( \frac{1}{\varrho^2} \!\cdot\! \frac{p^2}{M_X^2} \right)
\label{vertexnew}
\en
instead of the conventional one given in Eq. (\ref{vertex}).

In doing so, we keep the central values of the basic CCQM parameters
represented in Eq.~(\ref{centralparameters}). Namely, we use the universal
infrared cutoff parameter $\lambda = 0.181$ GeV and the constituent
charm quark mass in the range of $\pm 10\%$ around $m_c =1.67$ GeV.

For the numerical evaluation on the charmonium states we vary the fully
adjustable parameter $\varrho > 0$ to fit the latest experimental data 
on radiative charmonium decays reported by the PDG~\cite{PDG20}.

\begin{figure}[H]
\begin{center}
\includegraphics[scale=0.35]{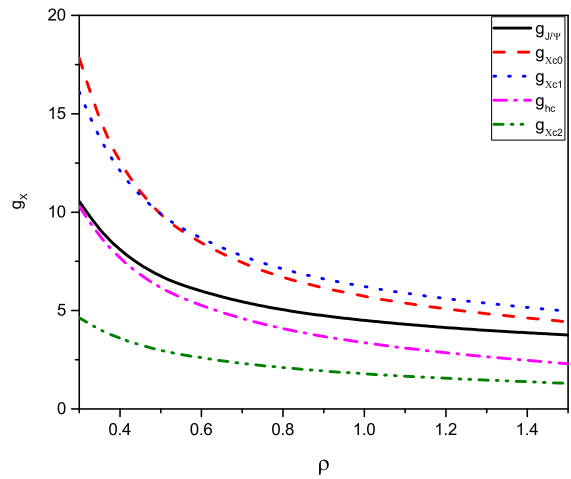}
\end{center}
\caption{
The renormalized couplings of the charmonium ground-states and first
orbital excitations calculated in dependence on the slope parameter
$\varrho$.
}
\label{fig3}
\end{figure}

{\bf 1}. First we calculate the renormalization couplings $g_H$ of the hadrons
which play an important role in the CCQM by excluding the constituent degrees
of freedom from the space of physical states. They are strictly fixed by the
compositeness requirements expressed in Eq.~(\ref{gren}) and do not constitute
further free parameters, although keep indirect dependencies on the basic model
parameters.
The renormalized couplings of the charmonium ground-states and first orbital
excitations calculated in dependence on the ``slope'' parameter $\varrho$ for
$m_c = 1.80$ GeV are shown in Fig.~\ref{fig3}. The renormalization couplings
$g_X$ decrease monotonically, a larger slope for $\varrho \leq 1$ and
more moderately for larger
values.

\begin{figure}[tb]
\begin{center}
\includegraphics[scale=0.35]{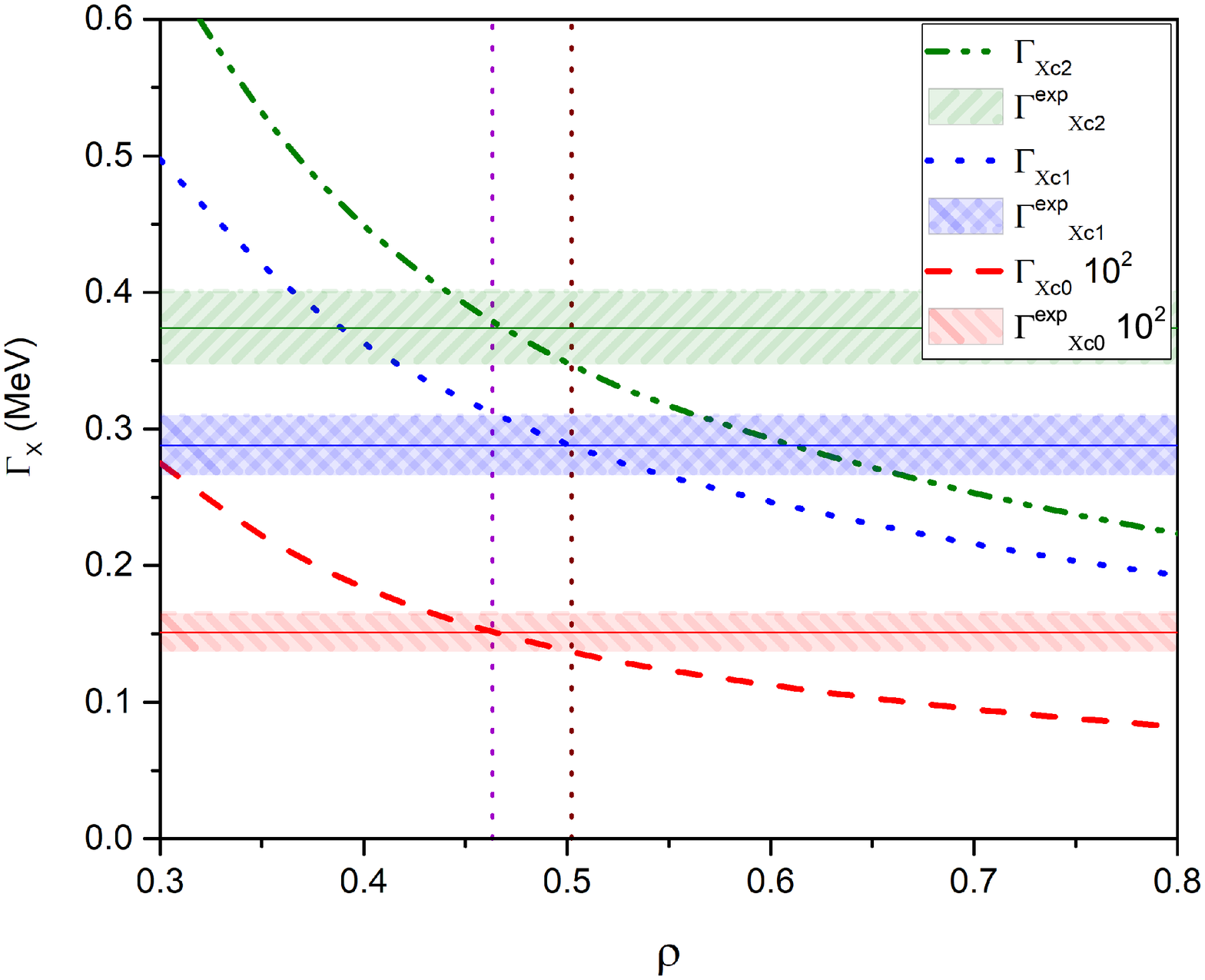}
\end{center}
\caption{
Partial decay widths of the dominant one-photon radiative decays of
the orbitally excited ($\emph l = 1$) charmonium states
$\chi_{cJ},~J=\{0,1,2\}$ in dependence on the slope parameter $\varrho$
for a fixed c-quark mass $m_c=1.80$~GeV. The shaded bands with
central horizontal solid lines correspond to the latest experimental data with
given error bars.
}
\label{fig4}
\end{figure}

{\bf 2}. Having calculated the renormalization couplings $g_X$ we are able to
estimate the partial widths of the dominant one-photon radiative decays of the
orbitally excited ($\emph l = 1$) charmonium states $\chi_{cJ},~(J=\{0,1,2\})$
by using Eqs.~(\ref{widthXc0}), (\ref{widthXc1}) and~(\ref{widthXc2}) to find the
optimal model value of the ``slope'' parameter $\varrho >0$.
The dependencies of the partial widths $\chi_{c0}$,
$\chi_{c1}$ and $\chi_{c2}$ on the ``slope'' parameter $\varrho$ are shown in
Fig.~\ref{fig4} for a fixed value of the charm quark mass $m_c=1.80$ GeV.
We conclude that our theoretical estimates fit the corresponding experimental
data only in the narrow interval $\varrho \in [0.47 \div 0.50 ]$.

\begin{figure}[H]
\begin{center}
\includegraphics[scale=0.4]{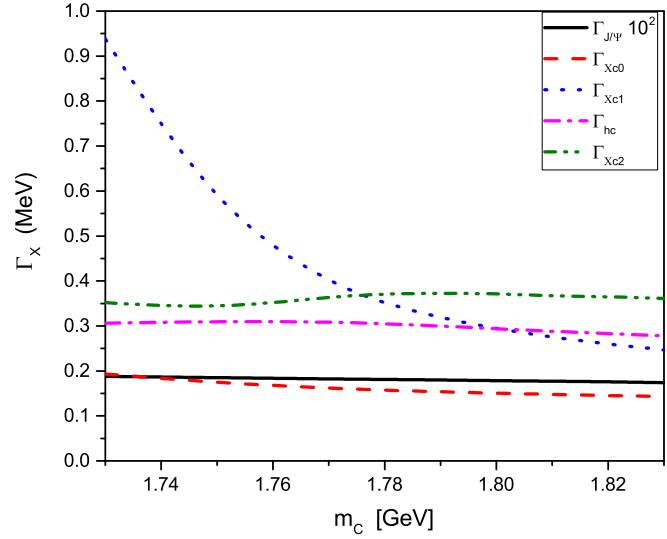}
\end{center}
\caption{
Theoretical estimates of the charmonium partial decay widths $\Gamma_X$ in
dependence on the charm quark mass $m_c$ for fixed values of the slope 
parameter $\varrho=0.485$.
}
\label{fig5}
\end{figure}

{\bf 3}. The dependencies of the charmonium partial decay widths $\Gamma_X$
on the charm quark mass $m_c$ for fixed values of the ``slope'' parameter
$\varrho=0.485$ are depicted in Fig.~\ref{fig5}. The best fits of our theoretical
estimates to the corresponding experimental data occur around $m_c=1.80$ GeV.

\begin{figure}[H]
\begin{center}
\includegraphics[scale=0.4]{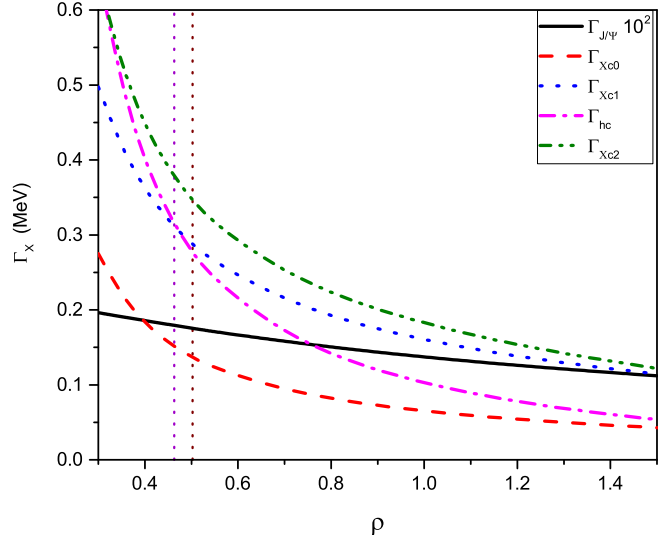}
\end{center}
\caption{
Partial decay widths of the dominant one-photon radiative decays of
the charmonium ground $J/\psi$ and orbitally excited ($\emph l = 1$) states
$\chi_{cJ},~J=\{0,1,2\}, h_c$ in dependence on the slope parameter
$\varrho$ for fixed c-quark mass $m_c=1.80$~GeV.
The vertical dotted lines indicate the optimal $\varrho$-interval from
the fit of $\chi_{cJ}$ one-photon decays.
}
\label{fig6}
\end{figure}

Having fixed the model parameter $m_c=1.80$ GeV  we are able to calculate
the partial widths of the dominant one-photon radiative decays of the ground
($J/\psi \to \gamma\eta_c$) and orbitally excited ($h_c \to \gamma J\psi$)
in dependence on $\varrho$, shown in Fig.~\ref{fig6} together with the curves
curves for $\chi_{cJ},~J=\{0,1,2\}$.

Finally by fitting the latest experimental data~\cite{PDG20} on the partial
widths of the dominant one-photon radiative decay of the
orbitally excited charmonium states $\chi_{c0}$, $\chi_{c1}$ and $\chi_{c2}$ we fix
the optimal values of model parameters as follows:
\begin{equation}
\begin{tabular}{ c c c }
\quad $\lambda$ \quad &
\quad $m_c$ \quad &
\quad $\varrho$ \quad \\
\hline
\quad 0.181 \ {\rm GeV} \quad &
\quad 1.80  \ {\rm GeV} \quad &
\quad 0.485
\end{tabular}
\label{fixedparameters}
\end{equation}

Note, the adjusted value of the 'slope' parameter $\varrho$ is common
to all charmonium states under consideration and is about 1/2. Subsequently, the
conventional 'size' parameters of these charmonia would be nearly half
of their physical masses. The refitted constituent c-quark mass value in
Eq.~(\ref{fixedparameters}) exceeds the central basic value in
Eq.~(\ref{centralparameters}) within the allowed $\sim 8\%$ deviation.

Our theoretical values for the fractional widths of the dominant one-photon
radiative decay of the S- and P-wave charmonia in comparison with some
recent theoretical predictions~\cite{brus2020,Becirevic:2012dc,weijun2016}
and experimental data~\cite{PDG20} are shown in Table~\ref{tab2}.

Note, the full decay width of the charmonium state $h_c({^{1}}\!P_{1})(3525)$ shown
in Table~\ref{tab1} is experimentally measured but with a large uncertainty (more than
$\pm 50 \%$) while the partial decay width of the radiative decay is detected
more accurately.

For the experimental value of $\Gamma(h_c \to \gamma \eta_c)$ given
in Table~\ref{tab2} we have used the latest data for the full decay width of $h_c$
and the corresponding fraction $\Gamma_i /\Gamma$ reported in~\cite{PDG20}
and shown in Table~\ref{tab1}.

\begin{table}[H]
\centering
\caption{
Some theoretical predictions of the partial widths (in units of keV) of the
dominant radiative decay of the charmonium states below the $D{\bar D}$
threshold in comparison with recent data.
}
\label{tab2}
\begin{tabular}{|c|c|c|c|c|c|c|c|c|}
\hline
$J^{PC}$ &
Radiative decay &
${}^{CCQM}_{\lambda=0.181}$ &
${}^{CCQM}_{~~\lambda \to 0}$ &
Exp~.~\cite{PDG20}  &
{p/m}~\cite{brus2020} &
{LWL}~\cite{brus2020} &
\cite{Becirevic:2012dc} &
\cite{weijun2016} \\
\hline
$1^{--}$ &
$\Gamma(J/\psi \to \gamma \eta_c$) &
1.771 &
1.771 &
1.58 $\pm$ 0.43 &
   -       &
   -       &
 2.64(11)  &
 1.25        \\
\hline
$0^{++}$ &
$\Gamma(\chi_{c0} \to \gamma J/\psi)$ &
142.0 &
142.0 &
151 $\pm$ 14 &
118 &
128 &
 - &
 128 \\
\hline
$1^{++}$ &
$\Gamma(\chi_{c1} \to \gamma J/\psi)$ &
296.7 &
297.0 &
288 $\pm$ 22 &
315 &
266 &
 -  &
 275 \\
\hline
$1^{+-}$ &
$\Gamma(h_c \to \gamma \eta_c)$ &
290.8 &
290.7 &
357 $\pm$ 270 &
   -      &
   -      &
 720(50)(20) &
 587 \\
\hline
$2^{++}$ &
$\Gamma(\chi_{c2} \to \gamma J/\psi)$ &
358.1 &
356.7 &
374 $\pm$ 27 &
419 &
353 &
 -  &
 467 \\
\hline
\end{tabular}
\end{table}

As mentioned in Sec. II, the uncertainties of our basic model parameters,
estimated renormalization couplings and form factors are about $10\%$,
while the branching fractions are proportional to the square of these values.
We therefore expect that the uncertainties of our calculation of the
partial widths in Table~\ref{tab2} do not not exceed more than  a few percent.

Concluding this section, we have fixed our model parameters $m_c$ and
$\varrho$ by fitting the latest data in~\cite{PDG20} on the partial decay widths
of the one-photon radiative decays of the orbitally excited charmonium
states $\chi_{c0}$, $\chi_{c1}$ and $\chi_{c2}$. By using these fixed
parameters, we have calculated
$\Gamma(J/\psi \to \gamma \eta_c)$ and $\Gamma(h_c \to \gamma \eta_c)$.

\section{Deconfinement limit $\lambda\to 0$}
\label{sec:deconfinement}

The infrared cutoff parameter introduced in the CCQM plays an important
role [see, e.g., Eq.~(\ref{integ1})]. It leads to the removal of possible threshold
singularities
corresponding to the creation of free quarks, and is taken to be universal
($\lambda=0.181$~GeV) for all physical processes. 
However, in some specific cases these singularities do not appear.
Particularly, the fixed value ($m_c = 1.80$ GeV) of the constituent charm
quark mass obeys the condition $M_X < 2m_c$ for all charmonium states
under consideration and the initial integral in Eq.~(\ref{integ0}) converges
and we can evaluate it for the full integration range $t\in[0,\infty)$ with
$\lambda \to 0$  in Eq.~(\ref{integ1}).

\begin{figure}[hb]
\begin{center}
\includegraphics[scale=0.3]{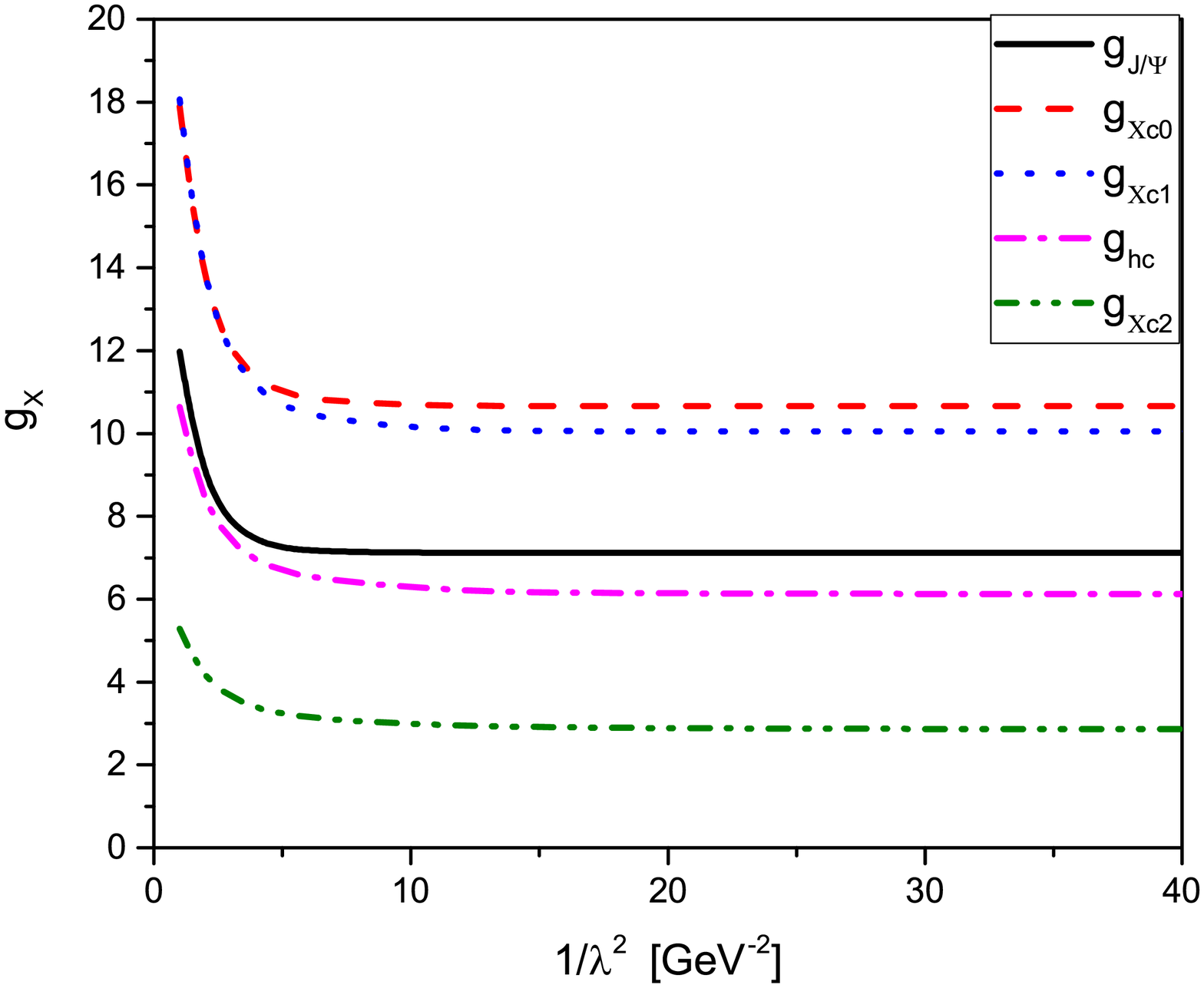}
\hspace*{-15mm}
\includegraphics[scale=0.3]{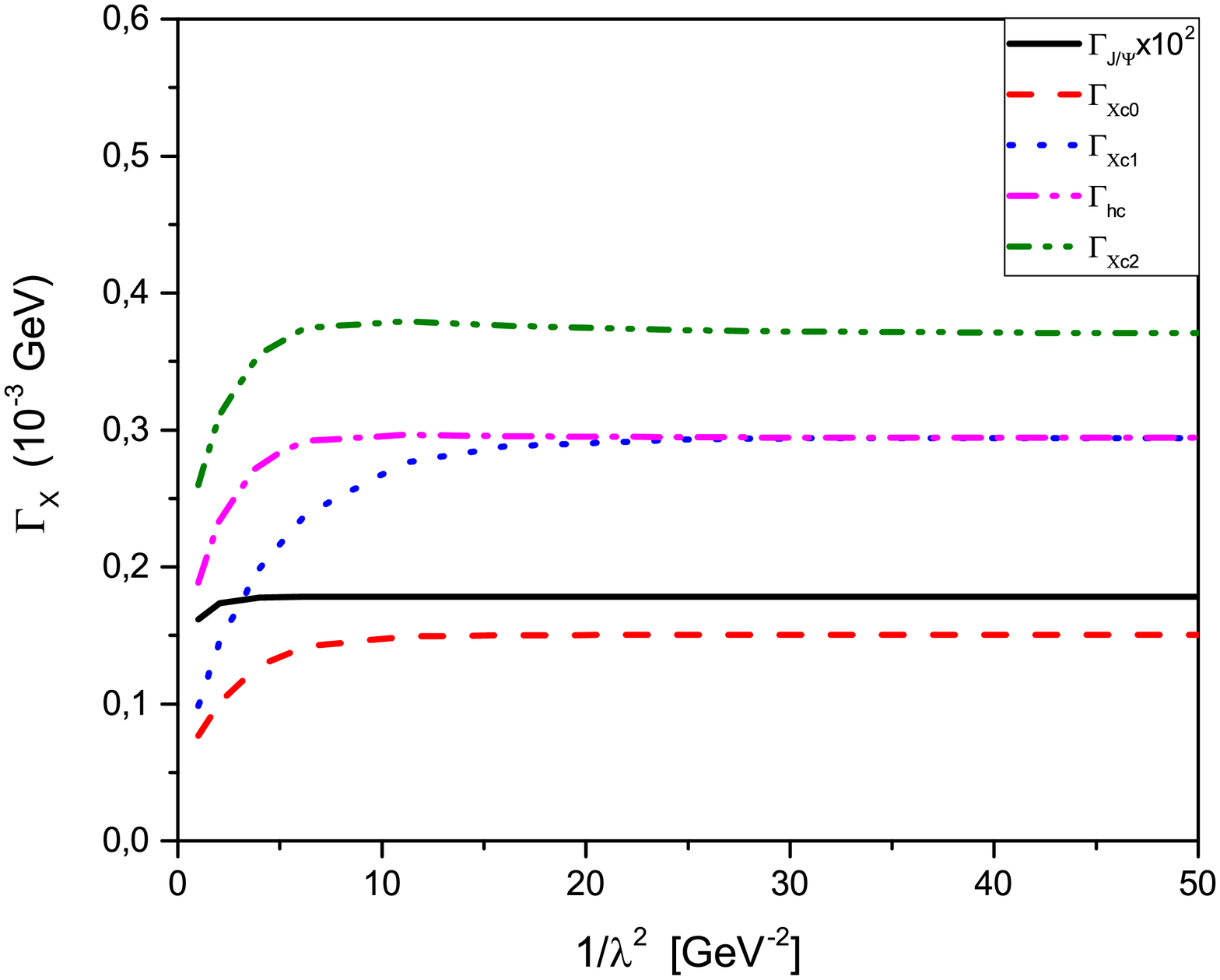}
\end{center}
\caption{
Dependence of the renormalized couplings $g_X$ and the one-photon radiative
partial decay widths $\Gamma_X$ of the charmonium states on the infrared
cutoff parameter $1/\lambda^2$.
}
\label{fig7}
\end{figure}

The dependencies of the renormalized couplings $g_X$ and the one-photon
radiative partial decay widths $\Gamma_X$ of the charmonium states on the
upper integration bound $1/\lambda^2$ is presented in Fig.~\ref{fig7}.
The renormalized couplings and partial decay widths depicted in Fig.~\ref{fig7}
do not change for $1/\lambda^2 > 20$ GeV$^{-2}$ while the CCQM universal
confinement parameter $\lambda=0.181$~GeV corresponds to
$1/\lambda^2 = 30.52$ GeV$^{-2}$.
Therefore, our theoretical estimates for the renormalization couplings and
fractional widths of the dominant one-photon radiative decay of the charmonium
states shown in Table~\ref{tab2} remain unchanged in the deconfinement
limit $\lambda\to 0$.

\section{Discussion}
\label{sec:discussion}

\subsection{ $J/\psi({}^3\!S_1) \to \gamma \eta_c({}^1\!S_0)$}

Nowadays, discrepancies still exist between theoretical predictions and the
data on the transition $J/\psi\to\gamma\eta_c(1S)$.

Particularly, the nonrelativistic potential model~\cite{Swanson:2005} and
the Coulomb gauge approach~\cite{Guo:2014zva} give a partial width of
$\Gamma[J/\psi\rightarrow \gamma\eta_c]\simeq 2.9$~keV, almost twice
as large as the present average data
$\Gamma^{\rm exp}(J/\psi\to\gamma\eta_c)\simeq 1.58\pm 0.37$~keV~\cite{PDG20}.

Lattice QCD simulations for the hadronic matrix elements~\cite{Becirevic:2012dc}
relevant to the radiative $J/\psi\to \gamma\eta_c$  and $h_c\to \gamma\eta_c$
decays have been performed by using the twisted mass QCD action with light
dynamical quarks ($N_{\rm f}=2$) for several small lattice spacings. The authors
smoothly extrapolated the relevant form factors at four lattice spacings to the
continuum limit. For the radiative decay $J/\Psi \to \gamma\eta_c$ the lattice
simulation computes a decay rate
$\Gamma(J/\psi\to\gamma\eta_c) = 2.64(11)(3)$ keV~\cite{Becirevic:2012dc}
that is obviously larger than the recent data~\cite{PDG20}.

Recently, the radiative decays of the charmonium states $J/\psi$, $h_c(1P)$
and $\chi_{cJ}(1P)$ have been studied within a constituent quark model.
The predicted partial decay width
$\Gamma(J/\psi\to\gamma\eta_c) \simeq 1.25$ keV~\cite{weijun2016} is
significantly (about 22$\%$) lower than the central value of the present
data~\cite{PDG20}.

According to our calculation, the partial decay width of the transition
$J/\psi({}^3\!S_1)\to\gamma\eta_c({}^1\!S_0)$ in Eq.~(\ref{widthJpsi})
is proportional to $\alpha \cdot (M_{J/\psi} - M_{\eta_c})^3 \cdot |C(...)|^2$.
Hereby, we note that the form factor $C$ in Eq.~(\ref{formJpsi}) is of order
$m_c (1/m_c^2 + 1/M_{J/\psi}^2+1/M_{\eta_c}^2)$ and
$m_c \sim M_{J/\psi}/2 \sim M_{\eta_c}/2$. Obviously, the transition
rate of $J/\psi \to \gamma\eta_c(1S)$ is strongly suppressed by the factor
$(M_{J/\psi} - M_{\eta_c})^3/m_c^2$.

Our calculation within the CCQM for the partial decay width (see Table II)
\begin{eqnarray}
\Gamma(J/\psi \to \gamma \eta_c) = 1.77 \ {\rm keV}
\end{eqnarray}
slightly (about 12$\%$) exceeds the average value of the recent
data~\cite{PDG20}.

\subsection{$ h_c({^{1}}\!P_{1}) \to \gamma \eta_c ({^{1}}\!S_{0})  $}

The charmonium state $h_c(1P)$ has been discerned from the experimental
background only recently, the CLEO succeeded to isolate this
state~\cite{Rosner:2005ry}. Only a few decay modes of the $h_c$ are observed.
Due to its negative $C$-parity, the dominant decay mode is
$h_{c}(1P)\to \gamma\eta_c(1S)$ with a branching fraction of
$(51 \pm 6)\%$~\cite{Rosner:2005ry,Dobbs:2008ec,PDG20}.

There are some discrepancies between the different model predictions for
$\Gamma [h_{c}(1P)\to \gamma\eta_c]$.
A constituent quark model prediction for the  partial decay width~\cite{weijun2016}
$\Gamma(h_c(1P)\to\gamma\eta_c)\simeq 587$ keV is consistent
with data within its large uncertainties~\cite{PDG20} but much smaller than
the predictions from the relativistic quark model~\cite{Ebert:2003} and lattice
QCD~\cite{Dudek:2006ej}.

A recent lattice calculation gives a larger width of
$\Gamma(h_{c}\to\gamma\eta_{c})=720(50)(20)$~keV~\cite{Becirevic:2012dc}.
A similar result was also obtained with a light front quark model~\cite{Ke:2013zs}.
Both predictions are obviously much larger than the data~\cite{PDG20}.

Our calculation within the CCQM for the partial decay width reads (see Table II)
\eq
\Gamma(h_c \to \gamma \eta_c) = 0.291 \ {\rm MeV}\,.
\label{hc-partialwidth}
\en

The present world data for the full decay width of the charmonium state
$h_c({^{1}}\!P_{1})(3525)$ cannot be used to test the various predictions due to their
large uncertainties~\cite{PDG20}.
On the other hand,  the fractional width for the one-photon radiative decay
of $h_c({^{1}}\!P_{1})(3525)$ is detected more accurately~\cite{PDG20}.
Therefore, by combining the latest value for the fractional width of~\cite{PDG20} with our
estimate in Eq.~(\ref{hc-partialwidth})  we may calculate the 'theoretical
full decay width' for $h_c$ as follows:
\eq
\Gamma^{\rm theor}_{h_c}  \simeq ( 0.57 \pm 0.12 ) \ {\rm MeV}             \,.
\label{theoretical}
\en
Hereby, we admitted a relevant $\sim 10 \%$ uncertainty for
$\Gamma(h_c \to \gamma \eta_c)$.
Compared with data $\Gamma^{\rm exp}_{h_c} \simeq (0.7\pm 0.4)$~MeV~\cite{PDG20}, 
the prediction in Eq.~(\ref{theoretical}) is located in a
more narrow interval.

\subsection{Triplet $ \chi_{cJ}({^{3}}\!P_{J}) \to \gamma J/\psi({}^3\!S_1)$}

Recently, a Cornell potential model has been used to study electromagnetic
transitions $\chi_{cJ}(1P) \to \gamma J/\psi$ in charmonium \cite{brus2020}.
The absence of any momentum dependence in the potential allowed
for a complete factorization of the heavy quarkonium mass ($m_c$)
dependence in the calculation of the electromagnetic decay widths in the
$\sim p_c/m_c$ and long-wavelength (LWL) approximations. The calculated
decay widths  $\Gamma_{p/m}^{(II)}$ and $\Gamma_{LWL}^{(II)}$
in~\cite{brus2020} are shown in Table~\ref{tab2} in comparison with our
present results. Hereby,  the effective charm-quark mass parameter, at least
for the description of the low lying charmonium states, is constrained to be
around $1.840$ GeV.
Estimates of $\Gamma(\chi_{c0,c1,c2}(1P)\to\gamma J/\psi)$ reported 
in~\cite{weijun2016} with a constituent quark model differ from  the average
data~\cite{PDG20} (see Table~\ref{tab2}).
Some studies on the radiative transition properties of $\chi_{c0,c1}(1P)$
were carried out in Lattice QCD as well~\cite{Dudek:2006ej,Chen:2011kpa},
however, good descriptions are still not obtained due to some technical problems.

Nowadays, discrepancies still remain between the different model predictions
for the partial decay widths $\Gamma(\chi_{cJ}(1P)\to\gamma J/\psi)$ of the
charmonium triplet $\chi_{c0,c1,c2}$.

Our calculations for the central values of the partial decay widths (see Table II)
read
\begin{eqnarray}
\Gamma(\chi_{c0}(1P)\to\gamma J/\psi) \!\!&=&\!\! 142.0 \, {\rm keV}\,,
\nonumber\\
\Gamma(\chi_{c1}(1P)\to\gamma J/\psi) \!\!&=&\!\! 296.7 \, {\rm keV}\,,
\nonumber\\
\Gamma(\chi_{c2}(1P)\to\gamma J/\psi) \!\!&=&\!\! 358.1 \, {\rm keV}
\end{eqnarray}
and are close to the recent LHCb data ~\cite{PDG20}.

\section{Summary}
\label{sec:summary}

The dominant one-photon radiative transitions of the charmonium $S$- and
$P$-wave states have been studied within the CCQM. In doing so, the
transition amplitudes have been described by the leading-order triangle
diagrams. We neglect the subleading bubble diagrams since their small
contributions are comparable to the common model error ($\sim\pm 10 \%$).
The gauge invariant transition amplitudes have been expressed by using
either the conventional Lorentz structures, or the helicity amplitudes, where
it was effective.

We used the basic model parameters for the global infrared cutoff
($\lambda=0.181$~GeV) and the constituent c-quark mass ($m_c=1.80$~GeV
which deviates $\sim 7\%$ from the central model value). We additionally
introduced only one adjustable parameter $\varrho>0$ common to the
charmonium states $\eta_c({}^1\!S_0)$, $J/\psi({}^3\!S_1)$,
$\chi_{c0}({^{3}}\!P_{0})$, $\chi_{c1}({^{3}}\!P_{1})$,  $h_c({^{1}}\!P_{1})$, 
and $\chi_{c2}({^{3}}\!P_{2})$ to describe the quark distribution inside the
hadron. This dimensionless parameter indicates the ratio between the
charmonium size parameter and its physical mass.

The renormalization couplings $g_X$ of the charmonium states play an important
role in the CCQM by excluding the constituent degrees of freedom from the space
of physical states. They are strictly fixed by the compositeness requirements
($Z_H=0$) and further eliminated as free parameters, although keep indirect
dependencies on the basic model parameters. We have analyzed the behavior
of $g_X$ in dependence on $\varrho$. The couplings $g_X$  decrease fast for
$\varrho \leq 0.8$ and slowly for larger values.

The optimal value of the only adjustable parameter $\varrho$ has been found
by fitting the latest data for the partial widths of the one-photon radiative
decays of the charmonium triplet $\chi_{cJ}({^{3}}\!P_{J}),~J=\{0,1,2\}$.  Note,
the optimal value $\varrho=0.485$ is close to one half and common to all
charmonium states under consideration. Then, we calculated fractional widths
for the states $J/\psi({}^3\!S_1)$ and $h_c({^{1}}\!P_{1})$. The results are
in agreement with the latest data.

By using the latest partial decay data from PDG2020~\cite{PDG20} and our estimated
partial decay width for $h_c({^{1}}\!P_{1})$ we calculate reversely the
theoretical 'full decay width' $\Gamma^{\rm theor}_{h_c}\simeq ( 0.57 \pm 0.12 )$
MeV that may be compared to
$\Gamma^{\rm exp}_{h_c} \simeq (0.7\pm 0.4)$ MeV~\cite{PDG20}.

The infrared cutoff parameter $\lambda$ plays the central role in the CCQM by
removing possible threshold singularities corresponding to the creation of free
quarks, and is taken to be universal ($\lambda=0.181$~GeV) for all physical
processes. However, the fixed charm-quark mass value ($m_c=1.80$~GeV)
excludes any appearance of branch points due to the mass relations
$M_X < 2m_c$ valid for charmonium states under consideration.
This allowed us to repeat our calculations by decreasing  $\lambda$ gradually
close to the deconfinement limit $\lambda\to 0$. Our results obtained
for $\lambda=0.181$~GeV changed insignificantly and converged to their limits
for  $1/\lambda^2 > 20$ GeV$^{-2}$ or $\lambda < 0.220$ GeV. For this
charmonium states our theoretical calculations performed within the CCQM
remain unchanged even in the infrared deconfinement limit.

\begin{acknowledgments}

This paper is devoted to the memory of our friend and colleague J\"urgen K\"orner. 
G.G. gratefully acknowledges support from 
the Alexander von Humboldt Foundation and would 
like to thank Institut f\"ur Theoretische Physik,
Universit\"at T\"ubingen for warm hospitality. 
M.A.I.\ acknowledges the support from the PRISMA Cluster of Excellence
(Mainz Uni.). 
This work was funded by BMBF
``Verbundprojekt 05P2018 - Ausbau von ALICE                                    
am LHC: Jets und partonische Struktur von Kernen''
(F\"orderkennzeichen No. 05P18VTCA1), by ANID (Chile) under
Grant No. 7912010025, by ANID PIA/APOYO AFB180002 (Chile),
by FONDECYT (Chile) under Grant No. 1191103,
by Millennium Institute for Subatomic Physics
at the High-Energy Frontier (SAPHIR) of ANID, Code: ICN2019\_044 (Chile). 

\end{acknowledgments}

\end{document}